\documentclass{ar2e_cecil}
\usepackage[dvips]{graphicx}
\jname{Annu. Rev. Astron. Astrophys.}
\jyear{2005}
\jvol{43}
\ARinfo{1056-8700/97/0610-00}

\newcommand\arcsec{$^{\prime\prime}$}
\newcommand\farcs{\hbox{$.\!\!^{\prime\prime}$}}
\newcommand\apj{ApJ }
\newcommand\apjl{ApJ }
\newcommand\apjs{ApJS }
\newcommand\aap{A\&A }
\newcommand\aaps{A\&AS }
\newcommand\aj{AJ }
\newcommand\mnras{MNRAS }
\newcommand\nature{Nature }
\newcommand\araa{ARA\&A }

\newcommand\pasj{PASJ }

\newcommand\rmp{RevModPhys }
\newcommand\rmxac{RMxAC }
\newcommand\lesssim{\mathrel{\hbox{\rlap{\hbox{\lower4pt\hbox{$\sim$}}}\hbox{$<$}}}}
\newcommand\gtrsim{\mathrel{\hbox{\rlap{\hbox{\lower4pt\hbox{$\sim$}}}\hbox{$>$}}}}
\newcommand\la{\lesssim}
\newcommand\ga{\gtrsim}

\begin{document}

\title{Galactic Winds}

\author{Sylvain Veilleux\affiliation{Department of Astronomy, 
University of Maryland, College Park, MD 20742; 
E-mail: veilleux@astro.umd.edu}\\
Gerald Cecil\affiliation{Department of Physics and Astronomy, 
University of North Carolina, Chapel Hill, NC 27599-3255; 
E-mail: cecil@physics.unc.edu}\\
Joss Bland-Hawthorn\affiliation{Anglo-Australian Observatory, 
Epping, NSW, Australia; E-mail: jbh@aaoepp.aao.gov.au}}

\begin{keywords}
galaxies: evolution --- galaxies: halos --- galaxies:
intergalactic medium --- galaxies: kinematics and dynamics --- galaxies: nuclei
\end{keywords}

\begin{abstract}
Galactic winds are the primary mechanism by which energy and metals
are recycled in galaxies and are deposited into the intergalactic
medium.  New observations are revealing the ubiquity of this process,
particularly at high redshift.  We describe the physics behind these
winds, discuss the observational evidence for them in nearby
star-forming and active galaxies and in the high-redshift universe,
and consider the implications of energetic winds for the formation and
evolution of galaxies and the intergalactic medium. To inspire future
research, we conclude with a set of observational and theoretical
challenges.

\end{abstract}

\maketitle

\section{INTRODUCTION}

\subsection{Fundamental role}

On the largest physical scales, the Universe is shaped by gravity and
the dark energy. On galactic scales, gravity is responsible for only
part of what we see through telescopes.  This has become clear from
the most comprehensive numerical simulations involving up to $10^{10}$
particles, which purport to track the evolution of individual galaxies
to the present day.
Although Cold Dark Matter (CDM) simulations account well for the observed
large-scale structure, problems emerge in regions of high density and
high density contrast. Even the most detailed simulations do not
explain adequately what we know of galaxies.

Contemporary astrophysics is entering a new era where a proper
understanding of galaxy formation and evolution forces us to confront
details of physical processes involved in feedback, including star
formation and evolution, energetic and chemical recycling in the
interstellar and intergalactic media (ISM, IGM), gas dynamics, and
ultimately plasma magnetohydrodynamics. Progress is complicated by:
(i) the need for comprehensive data over the full electromagnetic
spectrum at comparable sensitivity and spatial resolution; and (ii) a
lack of detailed theoretical and numerical models that can accommodate
the multi-wavelength observations.

In this review, we confront arguably the dominant feedback in galaxy
formation and evolution --- galactic winds (GWs). Also important is
radiative feedback. But even here, outflows likely play a key role in
clearing a path for the escaping radiation (e.g., Dove, Shull, \&
Ferrara 2000).

\subsection{Early History}

More than forty years ago, Lynds \& Sandage (1963) announced `evidence
for an explosion in the center of the galaxy M82.' In the following
year, Burbidge, Burbidge, \& Rubin (1964) remarked that `the activity
in M82 is yet another manifestation of the generation of vast fluxes
of energy by processes which are not yet properly understood.' This
comment echoed the discovery of the first quasar in 1962, and the slow
realization that many extragalactic radio sources must be enormously
energetic, powered by processes that could only be guessed at (Hoyle
\& Fowler 1963). However, these seminal papers note important
similarities between M82 and the Crab nebula when comparing their
optical and radio properties, and they imply a possible role for nuclear
star clusters in driving the central explosion. By the end of the
decade, black holes (BHs) were identified as the driving mechanism in
quasars and powerful radio sources (Lynden-Bell 1969).

Before the discovery of a central explosion in M82, few workers
discussed galactic outflows.  The observed corona and radio halo in
the Galaxy (Spitzer 1956; Baldwin 1955) motivated Burbidge \& Hoyle
(1963) to consider whether such a halo was in fact bound to the
Galaxy.  In 1957, van Woerden, Rougoor \& Oort provided the first
evidence of outflowing gas from the Galactic Center, and identified an
`expanding arm.' Starting with Moore \& Spiegel (1968), many authors
went on to consider the possibility of a central explosion to power
the radial outflow (e.g., van der Kruit 1971; Sanders \& Prenderghast
1974; Oort 1977).

In a parallel development, after the detection of broad [O~II]
$\lambda$3727 emission in some elliptical galaxies by Osterbrock
(1960), Burke (1968) suggested the presence of a galaxy-scale
wind. Interest in GWs continued and models grew in sophistication,
motivated largely by the observation that ellipticals have very little
ISM and therefore may have been swept clean by galaxy-scale outflows
(Johnson \& Axford 1971; Mathews \& Baker 1971).

Over the past 40 years almost the entire electromagnetic spectrum has
become accessible, and we have become aware of the ubiquity of
GWs. Their study is now a key area of research, and advances in
ground-based and spaceborne instrumentation have produced a flurry of
high-precision data on spatially resolved winds in nearby galaxies.
At low redshift, winds in gas-rich galaxies are studied because the
dense ISM accentuates the outflow.  The most recent development is the
discovery of powerful winds in high-redshift galaxies. These new
observations strongly implicate wind-related feedback processes as key
to the chemical and thermal evolution of galaxies and the IGM.

\subsection{Review structure}

Despite the importance of this topic, there has never been a
comprehensive review of GWs published in the {\em Annual Reviews of
Astronomy and Astrophysics}, although
articles on closely related topics were written by Tenorio-Tagle \&
Bodenheimer (1988) on superbubbles and Spitzer (1990) on the hot ISM.

It is only recently that astronomers have been able to compile
multiwavelength data on a significant sample of galactic winds.  We
recognize the need for these observations to be discussed within a
universal framework, where hydrodynamical CDM simulations are able to 
predict the onset of galactic winds and track their evolution in 
different environments.  However, numerical models that give repeatable
results at different resolution scales (e.g. mass) have only recently 
become possible and the published results are limited at the present
time (Springel \& Hernquist 2003).  

The principal goals of our review are to use both observational
constraints and theoretical predictions to discuss critically the
properties of GWs in the local and distant universe, and to evaluate
the importance of these winds for the formation and evolution of
galaxies and the IGM.  We focus on material published within the past
15 years.  Our discussion centers on three fundamental issues:

\begin{itemize}
\item {\em Their Nature.} What are they? Which conditions trigger
them?  What powers them: star clusters or BHs? Are they nuclear or
disk-wide? How energetic are they $-$ are they relatively quiescent or
explosive?  What mass, momentum, energy, and metals do they transport?
How far?

\item {\em Their Frequency of Occurrence.} How common were they in the
past and are they now? What is their duty cycle?  When did they
begin to blow?

\item {\em Their Impact.} How important are they?  What impact do they
have on the nucleus, bulge, disk, halo, and dark matter of the host
galaxy?  Are they the dominant source of feedback in galaxy evolution?
How do they influence the intergalactic environment? What are their
fossil signatures?
\end{itemize}

Section 2 outlines the basic physics behind the GW phenomenon, and
presents several useful formulae and prescriptions to help with data
interpretation.  Sections 3, 4 and 5 provide an observational summary
of wind-blown events in local star-forming and active galaxies.
Section 6 summarizes the results from surveys at high redshifts, and
\S 7 discusses the impact of winds on the galactic and intergalactic
environments. In the last section (\S 8), we raise some important,
unanswered questions and suggest future directions for research.

\section{BASIC PHYSICS}
  
\subsection{Sources of Energy}

The central question of what powers a GW dates back forty years to the
discovery of the wind in M82. For reasons to be discussed in \S5.1,
remarkably few winds are unambiguously AGN (active galactic nucleus)
or starburst driven.  Hence, we now summarize the basic physics behind
both kinds of energy sources.  \medskip
    
\subsubsection{STELLAR WINDS \& SUPERNOVAE}
   
Starburst winds are driven by the mechanical energy and momentum from
stellar winds and supernovae (SNe).  Models that synthesize the
evolution of populations of massive stars have estimated the mass and
energy returned from starbursts (e.g., Leitherer, Robert, \& Drissen
1992;
Leitherer \& Heckman 1995; Leitherer et al. 1999, {\em Starburst99}).
This technique combines stellar evolutionary models, atmospheric
models, and empirical spectral libraries to determine the observable
properties of the starbursting population. A wide parameter space is
explored by varying the shape and mass limits of the stellar initial
mass function (IMF), the metallicity, and the star formation rate
(SFR) history (e.g., instantaneous burst or constant SFR).  Stellar
evolutionary models have reproduced successfully most properties of
Wolf-Rayet (WR) stars, but overpredict WR mass-loss rates.  In
general, combined empirical results and theoretical predictions are
used instead of model-derived mass-loss rates (Leitherer et al. 1992).
For an instantaneous starburst, winds from OB stars dominate early on
($\la$ 3 Myr); next come WR stars with mass-loss rates $\sim10$ times
higher ($\sim$ 3 -- 6 Myr); and finally, core-collapse Type II SNe
dominate until $\sim$ 40 Myr when the least massive ($\sim$ 8
M$_\odot$) ones explode.

Figures 107, 108, 113, and 114 from Leitherer et al. (1999) show the
time evolution of the mass-loss rates and mechanical luminosities
calculated using {\em Starburst99} for two different starburst
histories and various metallicities.  In the solar-metallicity case
where the mass-loss rate and mechanical luminosity are constant beyond
$\sim$ 40 Myr and scale with SFR,
\begin{equation}
\dot{M}_* = 0.26~(SFR / M_\odot~{\rm yr}^{-1})~~M_\odot\ {\rm yr}^{-1} 
\label{eqn1}
\end{equation}
\begin{equation}
\dot{E}_* = 7.0 \times 10^{41}~(SFR/ M_\odot~{\rm yr}^{-1})~~{\rm erg~s}^{-1}.
\label{eqn2}
\end{equation}
Because $\dot{p}_* = \dot{M}_* V_* = \sqrt{2 \dot{M}_* \dot{E}_*}$,
the rate of momentum injection from SNe and stellar winds into the
environment can be calculated:
\begin{equation}
\dot{p}_* = 5 \times 10^{33}~(SFR/ M_\odot~{\rm yr}^{-1})~~{\rm dyne}.
\label{eqn3}
\end{equation}
In these expressions, the supernova rate
$\approx~0.02~(SFR/1~M_\odot~{\rm yr}^{-1})~~{\rm yr}^{-1}$.

For comparison, the starburst's radiation pressure, $\tau L_{\rm bol}/c$, is
\begin{equation}
\dot{p}_{\rm rad} = 1.3 \times 10^{34}~\tau L_{\rm bol, 11}~~{\rm dyne},
\label{eqn4}
\end{equation}
where $L_{\rm bol, 11}$ is the bolometric luminosity of the starburst
in units of 10$^{11}$ L$_\odot$ and $\tau$ is the optical depth to
radiation. In luminous IR galaxies where $L_{\rm bol} \approx L_{\rm
IR} (8 - 1000 \mu{\rm m}) \ge 10^{11} L_\odot$ and
\begin{equation}
SFR  \approx 17~L_{\rm IR,11}~~M_\odot~{\rm yr}^{-1}
\label{eqn5}
\end{equation}
(Kennicutt 1998), the radiation pressure in the optically thick case
($\tau = 1$) can be comparable to the pressure from the stellar
ejecta.

In general, stellar winds are important only in young ($\la$ 10$^7$
yr) starbursts that form many high-mass ($\ga$ 60 M$_\odot$) stars in
a metal-rich (Z $>$ Z$_\odot$) environment. In any other situation, SN
explosions dominate the energetics of the ISM.  SN explosions usually
dominate by the time GWs develop, but stellar winds may be important
in superbubbles.  Because the star formation activity in a starburst
is highly correlated spatially, the ejecta from hot stars and SNe in
the starburst region quickly interact through shocks, and mix with the
surrounding gas to produce a cavity of hot, metal-enriched
material. The thermalization efficiency is the percentage of the
mechanical energy from the starburst that heats the gas.
Unfortunately, as we show in \S 3 and \S 4, this quantity is poorly
constrained by observations.  Hydro simulations (\S 2.4) often assume
it is 100\%, {\em i.e.}  none of the energy injected by the starburst
is radiated away.  In reality, this efficiency is set by the
environment, particularly by the gas density, and may be $\la$ 10\% in
the dense cores of powerful nuclear starbursts (e.g., Thornton et
al. 1998;
Silich, Tenorio-Tagle, \& Mu\~{n}oz-Tu\~{n}\'on 2003; Stevens \&
Hartwell 2003; Melioli \& de Gouveia Dal Pino 2004).  Conversely,
large SN rates may increase the porosity of the ISM, and hence
reduce radiative energy losses (e.g., Larson 1974; Cant\'o, Raga, \&
Rodriguez 2000; Wada \& Norman 2001).  Current data favor values $\ga$
10\% for the thermalization efficiency (see \S 4.5).

\subsubsection{AGN}

The ubiquity of supermassive BHs at the center of galaxies (e.g.,
Kormendy \& Gebhardt 2001) suggests that BH activity may also power
some galactic-scale outflows by accretion:
\begin{equation}
\dot{E} \simeq 10^{11}~(\epsilon/0.01)~\dot{M}_{\rm acc}~~L_\odot, 
\label{eqn6}
\end{equation}
where $\dot{M}_{\rm acc}$ is the mass accretion rate in M$_\odot$
yr$^{-1}$ and $\epsilon$, the energy conversion efficiency in rest
mass units.  The mass accretion rate ranges from $\la$ 0.001
M$_\odot$~yr$^{-1}$ for low-luminosity AGN, $\sim$ 1
M$_\odot$~yr$^{-1}$ for Seyfert galaxies, and $\sim$ 100
M$_\odot$~yr$^{-1}$ for quasars and powerful radio galaxies. The
parameter $\epsilon$ depends on BH spin and the boundary conditions
near the event horizon, and can reach $\epsilon \approx 0.4$ (e.g.,
Krolik 1999; Agol \& Krolik 2000). Much of the energy released during
accretion may be tapped to drive a GW; this can occur through several
processes.

Radiative processes may be important in luminous AGN such as Seyfert
galaxies and quasars. Radiation can push on the surrounding gas via
electron scattering, scattering and absorption on dust grains,
photoionization, or scattering in atomic resonance lines. The last one
is important for O-star winds
and is suspected to be responsible for fast ($\sim$ 0.03 $c$) outflows
in broad absorption-line quasars (BAL QSOs) (Crenshaw, Kraemer, \&
George 2003).
In terms of dynamics, the small opacity to electron scattering
($\kappa_{es} = 7 \times 10^{-25} x$ cm$^2$ per hydrogen atom, with
$x$ the ionization fraction) makes this process unimportant relative
to radiation pressure acting on dust grains (effective opacity
$\kappa_d \sim$ 10$^{-21}$ cm$^2$ per H atom), unless the material
under consideration is within the dust sublimation radius, $r_{\rm
  subl} \approx 1~\sqrt{L_{46}}$ pc ($L_{46}$ is the UV/soft X-ray
luminosity in units of 10$^{46}$ erg~s$^{-1}$), and the dust
temperature exceeds $T_{\rm subl} \approx 1200$ K. Dopita et al.
(2002) have shown that dust absorption dominates over photoelectric
absorption when $U \ga \alpha_{\rm B} / (c \kappa_d) \sim 0.01$, where
$\alpha_{\rm B}$ is the Case B recombination rate to excited states of
hydrogen (Osterbrock 1989) and $U$ is the ionization parameter, {\em
  i.e.} the ratio of photon to electron number density.  Radiation
pressure on dust is therefore probably unimportant in low-luminosity
AGN and low-ionization nuclear emission-line regions (LINERs) where $U
\la 10^{-4}$ (e.g., Ferland \& Netzer 1983; Veilleux \& Osterbrock
1987). However, it may dominate the dynamics of narrow emission-line
regions (NLRs) in Seyferts and quasars (see \S 5.3) and may be
responsible for the similarity of the ionization parameter in these
objects ($U \approx 0.01$; Dopita et al. 2002).

Radiative heating may also be dynamically important in luminous AGN.
Krolik, McKee, \& Tarter (1981) found that gas at $T \ga 10^{4.5}$ K,
when exposed to the hard power-law radiation of the AGN, undergoes
runaway heating if $U \ga 10 - 25$. The result is a gas phase at $T
\approx 10^6 - 10^7$ K in which Compton cooling balances inverse
Compton heating. The sound speed of this gas, $c_s = \sqrt{kT/\mu m_p}
\approx 300~(T/10^7~K)^{\frac{1}{2}}~{\rm km~s}^{-1}$, exceeds the
rotation velocity of typical galaxies. So, gas subject to runaway
heating may expel a wind from the host galaxy (Begelman 1985).  Given
the condition on $U$, this wind must originate near the nucleus, being
fed by gas from the accretion disk itself or nearby (Begelman, McKee,
\& Shields 1983;
Balsara \& Krolik 1993 and references therein). For Seyfert galaxies,
a substantial wind with $T_{\rm wind}$ $\approx$ 1 $\times$ 10$^6$ K
and $V_{\rm wind}$ $\approx$ 200 -- 500 km~s$^{-1}$ is driven off if
$L/L_{\rm Edd}$ $\ga$ 0.08, where $L_{\rm Edd} \approx 3 \times
10^{12} (M_{\rm BH}/10^8 M_\odot)~L_\odot$, the Eddington luminosity
(e.g., Balsara \& Krolik 1993). Unlike starburst-driven winds,
Compton-heated AGN winds are directed along the minor axis of the {\em
accretion} disk and are not necessarily perpendicular to the {\em
galactic} disk.

Magnetic fields in accretion disks almost certainly play a critical
role in powering and collimating AGN winds.  Numerical simulations
suggest that weakly magnetized accretion disks are subject to
magnetorotational instability and inevitably produce magnetically
active coronae (e.g., Miller \& Stone 2000). The coronae likely
generate outflows that are further boosted by centrifugal force
(Blandford \& Payne 1982; K\"onigl \& Kartje 1994). Hydromagnetically
boosted outflows may be at the origin of the highly collimated,
relativistic jets in powerful radio-loud galaxies (see reviews by
Zensus 1997 and Worrall \& Birkinshaw 2004).  These narrow beams
radiate inefficiently (Scheuer 1974), so most of their mechanical
energy must heat and agitate the host ISM and the
intracluster/intergalactic media (ICM/IGM).  The scale over which the
mechanical energy of the jets is deposited depends on several factors,
including the distribution of ISM near the AGN, the jet power, and how
collimated the outflow is at its source.  Jets in radio-quiet systems
are loosely collimated and barely relativistic (\S 5.2 and \S 5.3).

\subsection{Wind-blown bubbles}

Depositing mechanical energy by the starburst or the AGN
over-pressurizes a cavity of hot gas in the starburst or near the AGN
that reaches temperature
\begin{equation}
T = 0.4~\mu m_H~\dot{E}/k\dot{M},
\label{eqn7}
\end{equation}
where $\dot{E}$ is the fraction of the mechanical energy injection
rate that is thermalized and $\dot{M}$ is the rate at which the mass
is heated. For a starburst, equation (7) can be re-written using
equations (1) and (2) and assuming $\mu = 1.4$:
\begin{equation}
T \approx 3 \times 10^8~\xi~\Lambda^{-1}~~{\rm K},
\label{eqn8}
\end{equation}
where $\xi$ is the thermalization efficiency of the mechanical energy.
The parameter $\Lambda$ is the mass-loading factor, the ratio of the
total mass of heated gas to the mass that is directly ejected by SNe
and stellar winds or by the AGN. It accounts for the possibility that
some of the ISM is mixed with the stellar or AGN ejecta. Note that
this tenuous hot gas will be a poor X-ray (bremsstrahlung) emitter
unless $\xi/\Lambda \ll 1$. The central pressure of the cavity is
\begin{equation}
P_o \approx 0.118~\dot{E}^{\frac{1}{2}}~\dot{M}^{\frac{1}{2}}~R_*^{-2} = 0.08~\dot{p}~R_*^{-2}~~{\rm dyne\ cm}^{-2},
\label{eqn9}
\end{equation}
where $R_*$ is the radius of the injection zone (Chevalier \& Clegg
1985). $R_*$ is the radius of the star-forming region in the case of a
starburst-driven wind. We can rewrite equation (9) using equation (3)
in this case:

\begin{equation}
P_o/k \sim 3 \times 10^5~(SFR/M_\odot~{\rm yr}^{-1}) (R_*/{\rm kpc})^{-2}~~{\rm K~cm}^{-3}.
\label{eqn10}
\end{equation}

This pressure can significantly exceed the pressure of the undisturbed
ISM, hence driving the bubble outflow.  The hot cavity then evolves
like a stellar wind-blown bubble, whose dynamics were analyzed by
several authors (e.g., Castor, McCray, \& Weaver 1975;
Weaver et al. 1977;
Ostriker \& McKee 1988; Koo \& Mckee 1992a,b). As hot gas expands
through the sonic radius, it cools adiabatically (radiative cooling is
small in comparison).  Beyond the sonic radius, the wind drives a
shock into the surrounding ISM and starts to sweep it into a shell of
shocked gas.  As the bubble boundary expands, the dense shell
accumulates ISM and gradually slows the bubble's expansion to much
less than the wind velocity. This marks the end of the adiabatic
``free expansion'' phase, whose duration is set by the mechanical
luminosity of the starburst or AGN and the density of the ISM.  At
high mechanical luminosity ($\dot{E} > 10^{46}$ erg~s$^{-1}$) and in
gas-poor galaxies such as ellipticals ($n \sim 0.001 - 0.1$
cm$^{-3}$), free expansion may continue until the outer shock has
traversed much of the host galaxy (e.g., Schiano 1985). In contrast,
the free-wind phase in gas-rich starbursts and Seyfert galaxies is too
short to observe.

Bubble evolution in gas-rich systems is described by the self-similar
Taylor-Sedov solutions to a point-source explosion (blast wave) in a
homogeneous medium (Taylor 1950; Sedov 1959).
An extension by Schiano (1985) considered a continuous injection of
wind energy in the Kompaneets approximation of an exponential
atmosphere.  After free expansion, the system develops an
onion-like structure of five concentric zones; from smallest to
largest radii there is (1) the energy injection zone where the mass
and energy from the starburst or AGN is injected into the ISM, (2) a
free-flowing, supersonic outflow immediately outside the injection
zone, (3) a region of hot, shocked wind material, (4) a thin, dense
shell of shocked ISM, and (5) undisturbed ISM.
Note that the shocked ISM quickly becomes thermally unstable and
collapses into a thin shell.

If radiative losses of the overall system are negligible, the
expanding bubble is {\em energy-conserving}. In that case, the radius
and velocity of the expanding shell of shocked ISM are given by
(Castor et al. 1975; Weaver et al. 1977)
\begin{equation}
r_{\rm shell} = 1.1~(\xi\dot{E}_{44}/n_o)^{1/5}~t_6^{3/5}~~{\rm kpc},
\label{eqn12}
\end{equation}
\begin{equation}
V_{\rm shell} = 640~(\xi\dot{E}_{44}/n_o)^{1/5}~t_6^{-2/5} = 670~(\xi\dot{E}_{44}/n_o)^{1/3}~r_{\rm shell, kpc}^{-2/3}~~{\rm km~s^{-1}},
\label{eqn13}
\end{equation}
\begin{equation}
(\xi\dot{E}) = 3 \times 10^{35}~n_o~r_{\rm shell, kpc}^2~V_{\rm shell}^3~~{\rm erg~s}^{-1},
\label{eqn14}
\end{equation}
where $t_6$ is the age of the bubble in Myr, $n_o$ is the ambient
density in cm$^{-3}$, and $\dot{E}_{44}$ is the mechanical luminosity
of the wind in units of 10$^{44}$ erg~s$^{-1}$ (see, e.g., Shull \&
Saken 1995; Oey \& Massey 1995; Oey 1996 for more general cases).

{\em Momentum-conserving} bubbles --- where radiative losses are
significant --- decelerate faster: $r_{\rm shell} \propto
t^{\frac{1}{2}}; V_{\rm shell} \propto t^{-\frac{1}{2}} 
\propto r_{\rm shell}^{-1}$ (e.g., 
Steigman, Strittmatter, \& Williams 1975; Koo \& McKee 1992a, 1992b).
Their momentum injection rate is
\begin{equation}
\dot{p} \simeq 2 \times 10^{34}~n_o~r_{\rm shell, kpc}^2~V_{\rm shell, 100}^2~~~{\rm dyne}.
\label{eqn15}
\end{equation}

\subsection{Galactic winds}

Once the shell has formed, if the wind-blown bubble approaches the
scale height of the disk $H$, the shell reaccelerates, begins to
fragment through growing Rayleigh-Taylor (RT) instabilities, and
finally vents these fragments and the freely flowing and shocked wind
into the galaxy halo (e.g., Figure 1).  
Energy-driven GWs (we discuss momentum-driven
winds below) must be sufficiently energetic or long-lived to
``breakout''.  Radiative losses in the wind must be mild enough not to
drain too much energy from the wind to stall it; the cooling timescale
must exceed the time for the bubble radius (Equation [11]) to reach $\sim
H$.  This occurs in the hot GW fluid (MacLow \& McCray 1988), unless
$\xi/\Lambda \ll 1$ (Equation [8]).
The wind will break out if the expansion velocity of the shell
(Equation [12]) near $H$ exceeds the effective sound speed of the disk
gas. The approximate criterion is
\begin{equation}
\xi\dot{E} \ga 3 \times 10^{43}~~H_{\rm kpc}^{2}~P_7^{3/2}~n_o^{-\frac{1}{2}}~~~{\rm erg~s}^{-1},
\label{eqn16}
\end{equation}
where $H_{\rm kpc}$ is the disk scale height in kpc and $P_7$, the
ambient pressure $P/k$ in 
10$^7$ K cm$^{-3}$, typical of starbursts (Schiano 1985; MacLow \&
McCray 1988; Norman \& Ikeuchi 1989; MacLow, McCray, \& Norman 1989;
Koo \& McKee 1992a; Strickland et al. 2004b).

The steady-flow wind is a strong function of polar angle seen from the
starburst: The flow is a free wind up to the critical angle at which its
ram pressure becomes comparable to the thermal pressure in the diffuse
ISM. At larger angles, a standing bow shock in the galaxy disk
decelerates and deflects the wind around undisturbed ISM.  The
terminal velocity of the wind can be estimated by equating the total
energy deposition rate $\xi\dot{E}$ to the asymptotic rate of kinetic
energy (KE) loss: $\frac{1}{2}~\Lambda \dot{M} V_{\infty}^2 \approx
\xi \dot{E}$.
For a starburst-driven wind (eqns. 1 and 2) we obtain 
\begin{equation}
V_{\infty} \approx (2~\xi \dot{E}/\Lambda \dot{M})^{\frac{1}{2}} \approx 3000~(\xi/\Lambda)^{\frac{1}{2}}~~{\rm km~s}^{-1}.
\label{eqn18}
\end{equation}
The $\Lambda$ dependence is easy to understand: cold ISM gas that
feels the full brunt of the wind is shock heated and evaporated and
eventually mass loads the hot flow, which slows the wind.
Equation (16) assumes negligible halo drag.

Because momentum from the starburst or AGN cannot radiate away, a
momentum-driven wind may exist even when most of the energy is gone.
Then $\dot{p} = \Lambda \dot{M} V_{\infty}$,
where $\dot{p}$ is the sum of the radiation and mechanical momentum
from the starburst or AGN. Outflow occurs if this exceeds the gravity
of the host galaxy (e.g., Murray, Quataert, \& Thompson 2005).

\subsection{Hydrodynamical Simulations}

The main impediment to understanding the dynamical evolution of GWs is
that the bulk shocked wind is too hot to be studied by current X-ray
observatories.  Soft X-ray-emitting gas has been mapped (e.g.\ Fig.\ 2
\& 3), but is of uncertain origin, seems to fill only a few percent of
the volume, and contains at most 10\% of the wind mass and energy.
Simulations should therefore play a vital role in unraveling the
dynamical evolution of this and other gas phases involved in the flow.
However, gas at the shock velocities spanned by GWs ($\sim20$ to
$\ga2000$ km~s$^{-1}$; \S 4.4) is unstable both dynamically and
thermally: even several percent amplitude, pre-shock density
fluctuations (Innes, Giddings, \& Falle 1987a,b) are crushed by a
60-fold density increase while cooling to $10^4$ K.  Consequently,
many computational cells are required to resolve cooling filaments and
to track steep gradients.  In fact, all but the most recent
simulations have had to be two-dimensional, even though such
axisymmetry closes ``easy'' dimensions to entropy flow and thereby
forces disorder to emerge in inappropriate ways.  Even simulations
that adopted axisymmetry could handle only a limited range of gas
densities. It was necessary to model separately the collision of the
wind with low-density, diffuse ISM, and the collision with
higher-density clouds found near the galaxy disk plane.

Axisymmetric simulations have explored a range of disk/halo values
(e.g. Tomisaka \& Ikeuchi 1988; Tomisaka \& Bregman 1993), wind
luminosities (e.g. MacLow \& Ferrara 1999), mass loadings (Suchkov et
al 1996; Hartquist, Dyson, \& Williams 1997), metal versus mass-loss
rates
(especially in dwarf galaxies, e.g. MacLow \& Ferrara 1999; D'Ercole
\& Brighenti 1999), and mass and energy deposition histories in the
starburst (e.g. Strickland \& Stevens 2000).  Denser halos enhance
emissivity, and are posited either from a larger dark matter load, or
from high-latitude accretion debris that the wind may overrun.  A
disturbed, rarified, hot atmosphere from previous outbursts has been
incorporated into some AGN jet simulations (e.g. Smith et al. 1983),
but not yet for starburst winds.

\begin{figure} 
\caption{\small An example of a state-of-the-art 3D hydrodynamical
simulation of a starburst outflow in an ISM with fractal size
distribution at three epochs after the constant energy input wind
starts to blow: 1.0 Myr (bottom), 2.5 Myr (middle), and 4.0 Myr
(top). Blue is log-density and red is log-temperature (courtesy
J. Cooper, G. Bicknell, R. Sutherland).}
\end{figure}

After the wind impacts a dense cloud,
(Klein, McKee, \& Colella 1994; Silich et al. 1996; Hartquist, Dyson,
Williams 1997; Poludnenko, Frank, \& Blackman 2002; Williams \& Dyson
2002),
subsonic flow at the apex becomes transonic around the cloud edges,
inducing Kelvin-Helmholtz (KH) instabilities. Clouds therefore soon
develop a core/structured halo. Filaments in the halo disappear by
photo- or thermal-evaporation before moving far from the core.
Schiano, Christiansen, \& Knerr (1995, also Vietri, Ferrara, \&
Miniati 1997) show that the halo penetrates only to the depth of the
KH instabilities. The halo ablates from the unscathed core, dumping
the entropy induced by the enveloping shocked wind in a series of
``shedding events'' that ``fire polish'' the cloud surface to reset
the KH clock. By ablating small bumps, the cloud core stabilizes
itself against disruptive larger instabilities and can survive to
accelerate toward the wind speed.

In more energetic outflows into a denser ISM, mass loading can cause
RT instabilities to culminate in repeated, large vortices that crush
gas along shocks as the bubble apex shreds. Elsewhere the wind flow
can be almost adiabatic, progressing so rapidly that the timescales
for recombination, collisional ionization, and excitation exceed
greatly the dynamical time (Breitschwerdt \& Schmutzler 1999).
The gas soon becomes very convoluted, with a fractal size distribution
that Sutherland, Bicknell, \& Dopita (2003a) show enhances cooling,
hence shock prominence, compared to steady-flow models.

Steady winds can develop a ``stagnation ring shock'' in the disk plane.
Tenorio-Tagle \& Mu\~{n}oz-Tu\~{n}\'on (1997, 1998) show how this
shock becomes prominent when the ram pressure of even diffuse H I
augments the thermal pressure of the enveloping gas.  Wind fueling
stalls the ring at constant radius, a result consistent with
the sparse kinematical data on ring shocks (\S 4.3).
Feedback occurs because the base of the wind controls the minimum mass
accreted for the starburst to maintain the wind, but computer
limitations have thus far prevented coupling of starburst and
hydrodynamical codes of useful resolution.

In fact, most winds fail to develop a large-scale, steady flow.  The
small free-wind zone is then bounded at smallest radius by a few cells
of energy injection at the base, and at largest radius by the boundary
between dense clouds and their upstream bow shocks that appear
as elaborate filamentation at the resolution imposed by the grid.
Figures 2 and 3 show that the boundary emits most of the soft X rays.
Figure 1 is an example of a recent three-dimensional simulation that,
by detailing this region, promises to pin down the elusive X-ray
filling factor.

\section{MAJOR WINDBLOWN EVENTS IN THE LMC \& GALAXY}

The Galaxy and the Large Magellanic Cloud are excellent laboratories
for detailed studies of wind-blown events. The processes involved in
these local events are scaled down versions of superbubbles and GWs in
starburst galaxies. We summarize key features of these local
laboratories in this section.

\subsection{Large Magellanic Cloud}

The largest H~II region in the LMC (and indeed Local Group), 30
Doradus, is a microcosm of starburst processes.  Cluster R136 powers
the nested shells and superbubbles.  This mini-starburst contains
$\sim$50 very massive stars and has an estimated initial mass of a few
10$^4$ M$_\odot$ (Malamuth \& Heap 1994; Brandl et al. 1996; Brandl
2005). 
For comparison, M82 is powered by the equivalent of $\sim100$ R136's
within a region only 2-3 times larger than 30 Dor (Rieke et al. 1980;
Muxlow et al. 1994; O'Connell et al. 1995).
Shells and compact knots moving at $\sim$ 200 km~s$^{-1}$ are
detected in 30 Dor (Chu \& Kennicutt 1994; Redman et al. 2003),
perhaps forming the base of a large-scale wind able to escape the LMC
($v_{\rm esc} \approx 150$ km~s$^{-1}$).  Many shells in 30~Dor seem
to be momentum-conserving, not pressure-driven.
 
\subsection{Nuclear Wind in the Galaxy}

Only 8.0$\pm$0.5 kpc (Reid 1993) distant, the Galactic Center shows
remarkable energetic activity at infrared (IR), radio, X-ray and $\gamma$-ray
wavelengths (Morris \& Serabyn 1996; Yusef-Zadeh et al. 2000, 2005;
Cheng et al.  1997). While this activity has proved difficult to
disentangle, there is now solid evidence on scales of arcminutes to
tens of degrees for powerful mass ejections from the Galactic Center.
The idea of a central explosion dates back to the early discovery of
peculiar H~I kinematics there (see \S1.1), but we know today that
most, but not all, of the H~I kinematical signature is due to
streaming motions arising from a central bar (Morris \& Serabyn 1996).

A particular problem with GW studies has been deriving reliable
energies from multi-wavelength observations at comparable resolution.
Current estimates of the energetics of our Galactic Center span a huge
range.  Sofue \& Handa (1984) discovered the 200 pc diameter Galactic
Center radio lobe (GCL; Fig.\ 2$b$), with an implied thermal energy of
$\sim3\times 10^{51}$ erg. Chevalier (1992) argued for a higher value
($\sim 2\times 10^{52}$ erg) due to winds from hot young stars over
the past 30 Myr.  Others have argued from the high temperatures
implied by the ASCA detection of 6.7~keV K$\alpha$ emission from
He-like Fe~XXV (Koyama et al.\ 1989, 1996; Yamauchi et al. 1990; but
see Wang, Gotthelf, \& Lang 2002) that an explosive event ($4-8\times
10^{53}$ erg) must have occurred.  Bland-Hawthorn \& Cohen (2003)
detected the GCL at mid-IR wavelengths (Fig.\ 2$b$). The association
of the lobe with denser material raises the energetics to
$10^{54}/\kappa$ erg, where $\kappa$ is the covering fraction of the
dense shell; less energy is needed if there is substantial polycyclic
aromatic hydrocarbon (PAH) 
emission with the mid-IR continuum.  These energetics assume a shell
velocity of $\sim$ 150 km~s$^{-1}$, a value based on the kinematics of
the neighboring molecular gas after correction for bar streaming
(Bally et al. 1988); this value is uncertain because of our location
in the plane.
The ROSAT 1.5 keV diffuse X-ray map over the inner 45$^\circ$ 
provides compelling evidence for this GW interpretation (Fig.  2$a$)
(Bland-Hawthorn \& Cohen 2003).  Evidence for outflows on smaller
scale may be present in {\em Chandra X-ray Observatory} (CXO) X-ray
maps (Baganoff et al. 2003).

\begin{figure} 
\caption{\small Aspects of the Milky Way's wind. (top) ROSAT 1.5 keV
diffuse X-ray map that shows a biconical pattern emerging from the
Galactic Center on scales of tens of degrees.  (bottom) The inner
$\sim2.5\times2.5^\circ$ around the Galactic Center.  Above the plane
in red is the Galactic Center Lobe (GCL), here imaged by Law \&
Yusef-Zadeh with rasters from the Green Bank Telescope
(GBT). Elsewhere, the color image comes from 8.3 (B), $\sim13$ (G),
and 21.3 (R) $\mu$m scans obtained with the SPIRIT III radiometer on
the MSX spacecraft. These show embedded dust at various
temperatures. Note the warm dust filaments along the edges of the
GCL.}
\end{figure}

Potential energy sources are young star clusters or the $3-4\times
10^6$ M$_\odot$ central BH (Oort 1977; Frogel 1988; Genzel et al.
1996; Sch\"odel et al. 2003; Ghez et al.  1998, 2005). Individual star
clusters have ages ranging from 5~Myr (Krabbe et al.  1995) to 20~Myr
(Eckart, Ott, \& Genzel 1999; Figer et al.  2000).  While the star
formation history is undoubtedly complicated, there is now abundant
evidence that the Galactic Center has experienced several starburst
episodes (e.g., Tamblyn \& Rieke 1993; Sjouwerman et al.  1998;
Simpson et al.  1999).  Detailed models of PAH and fine structure
features (Lutz 1998) suggest a starburst $\sim$7~Myr ago, supported by
a census of stars (Genzel et al. 1994;
Krabbe et al.  1995; Najarro et al.  1997).

Activity seems to be fueled from the central molecular zone (CMZ), a
``ring" at 180~pc radius with M$_{\rm cmz} \sim 8\times 10^6$
M$_{\odot}$.  Inflow rates of $\sim1$ M$_\odot$ yr$^{-1}$ to the
Galactic Center (Morris \& Serabyn 1996) suffice to trigger starbursts
and nuclear activity in Seyfert galaxies ($\sim$10$^{43}$
erg~s$^{-1}$).  Hydro simulations (\S2.4) show that a central
explosion of $\sim10^{55-56}$ erg would provide mass M$_{\rm cmz}$
with sufficient radial impulse to make the observed ring (Sanders
1989; Saito 1990).

\section{LOCAL CENSUS OF GALACTIC WINDS IN STAR-FORMING GALAXIES}

In the next two sections, we present a census of galactic winds.
Each newly discovered outflow source provides different insights
into the wind process, and often reveals new and complex behavior.
Therefore, it is necessary to summarize what is known of the wind
phenomenon in terms of basic observational parameters (e.g.
morphology, kinematics, energetics). The subject of galactic winds
is largely in its infancy; therefore a discussion of the
phenomenology is warranted here.  We start by summarising the search
techniques used to find outflow sources, the limitations of each
technique and the inferred detection rates.

\subsection{Search Techniques \& Observational Limitations}

The wind phenomenon is likely to be sufficiently complex that a 
multiwavelength approach is necessary in order to probe all gas phases.
Line and continuum emission from warm,
hot, and relativistic plasma have been used with great success to
identify or confirm GWs.  Choice emission lines are H$\alpha$, [N~II]
$\lambda\lambda$6548, 6583, [S~II] $\lambda\lambda$6716, 6731, [O~III]
$\lambda$5007, and [O~II] $\lambda$3727 in the optical, and
Pa$\alpha$, Br$\gamma$, and H$_2$ 2.122 $\mu$m in the near-IR.
Ly$\alpha$ $\lambda$1216 should be avoided because resonant scattering
and selective dust absorption severely distort its emission profile
and intensity (e.g.,
Tenorio-Tagle et al. 1999; Keel 2005).  CXO and the XMM-Newton
Observatory have provided powerful new tools for complementary
searches in soft X rays, whereas facilities such as the upgraded {\em
  Very Large Array} (VLA) will continue efficient radio studies of
winds.

Enhanced emission along the minor axis of the host galaxy does not
necessarily imply a GW; other origins include tidal interaction and
ram-pressure stripping by the ICM.  Kinematic signatures of
extraplanar, non-gravitational effects are needed to be certain.
Double-peaked emission-line profiles attributable to an expanding,
shocked shell are often evident in wind galaxies. Emission-line ratios
typical of shock excitation (e.g., [N~II] $\lambda$6583/H$\alpha$
$\ge$ 1; [S~II] $\lambda\lambda$6716, 6731/H$\alpha$ $\ge$ 0.5)
indicate winds, but these diagnostics favor powerful ones with large
outflow velocities and a weak starburst radiation field (see \S4.8 for
more detail).

Selection effects hamper detection of GWs: ``Since deceleration is
what powers all of the emission process, the brightness distribution
due to a particular emission process will be dominated by the sites of
optimum local deceleration (corresponding to an optical density and
temperature contrast) within the range of decelerations that can give
rise to that process" (Braun 1985).  Observable winds are likely those
that are only slightly more pressured than the galaxy potential. Highly
energetic winds ($\gg 10^{58}$ erg or $\gg 10^{44}$ erg~s$^{-1}$ from
Equation [15]) may escape undetected, although this will depend on such
details as how energy is injected at the source, and the amount of gas
in and around the host galaxy.
Winds $\ll10^{41}$ erg~s$^{-1}$ would
have little observable impact on the ISM of an external galaxy.

The need to distinguish wind-related emission from the background
favors detection of highly inclined outflows in edge-on galaxies. This
is true for most of the well-known starburst-driven winds. In a few
less inclined galaxies (e.g., NGC~253, NGC~1808), outflows were first
suspected on the basis of absorbing dust filaments seen against the
galaxy body (e.g., Phillips 1993).  In NGC~253, extended H$\alpha$
filaments are projected onto disk H~II regions, requiring detailed 3D
kinematics to separate the different components. Similar blending
occurs at X-ray and radio wavelengths because compact sources are
commonly associated with the inner disk.  In edge-on systems,
kinematical deprojection leads to large uncertainties in wind
energetics. Other complications include uncertain dereddening at the
base of the outflow, relating the post-shock to the pre-shock
velocities, and high density bias introduced by the density-squared
dependence of the emission measure.

Absorption-line techniques have been used with great success to search
for the unambiguous wind signature of blueshifted absorbing material
in front of the continuum source. This method favors detection of
winds in face-on systems and therefore complements the emission
technique.  The equivalent width and profile of the absorption line
are used to estimate the amount of outflowing material along the line
of sight and to determine its projected kinematics. For an unsaturated
line, the column density of the absorbing material scales linearly
with the equivalent width of the line and therefore is arguably a
better probe of the whole range of density in the wind than the line
or continuum emission. For saturated lines, the change in equivalent
width reflects primarily a change in the velocity dispersion of the
absorbing clouds, covering fraction or both.

UV interstellar absorption lines are useful.  Low-ionization lines
such as
Si II $\lambda$1260, O I + Si II $\lambda$1303, C II $\lambda$1334,
Si II $\lambda$1526, Fe II $\lambda$1608, and Al II $\lambda$1670
are particularly well suited to avoid confusion with possible stellar
photospheric features of higher ionization (e.g., C III $\lambda$1176,
O IV $\lambda$1343, S V $\lambda$1501) and stellar wind features with
P Cygni-type profiles (e.g., N V $\lambda\lambda$1238, 1242, Si IV
$\lambda\lambda$1393, 1402, C IV $\lambda\lambda$1548, 1550, He II
$\lambda$1640). A few key optical lines have also been used in
successful wind searches in $z \la 0.5$ galaxies: Na~I~D
$\lambda\lambda$5890, 5896 and K I $\lambda\lambda$7665, 7699.
Extension to the far-UV domain with the {\em Far Ultraviolet
  Spectroscopic Explorer} (FUSE) spacecraft has recently allowed the
use of the O$\:$VI $\lambda\lambda$1032, 1038 lines to probe the
coronal-phase ($T$ = a few $\times$ 10$^5$ K) gas in GWs. Such
measurements constrain the amount of radiative cooling (see \S 4.5).

\subsection{Detection Rate}

There are now estimates of the frequency of wind occurrence in nearby
star-forming galaxies for most wavebands.  Detailed studies of
extended optical line emission around various starburst and quiescent
galaxies reveal a clear trend between starburst strength and the
presence of kinematically confirmed winds or extraplanar diffuse
ionized gas (e.g., Heckman, Armus, \& Miley 1990; Hunter, Hawley, \&
Gallagher 1993; Pildis, Bregman, \& Schombert 1994; Lehnert \& Heckman
1995, 1996; Marlowe et al. 1995;
Veilleux et al. 1995; Rand 1996; Hunter \& Gallagher 1997; Kim,
Veilleux, \& Sanders 1998; Martin 1998; Miller \& Veilleux 2003a,
2003b; Rossa \& Dettmar 2000, 2003a, 2003b; although see Meurer 2004).
Complementary results from systematic searches for blueshifted optical
absorption lines (e.g., Heckman et al. 2000; Rupke, Veilleux, \&
Sanders 2002, 2005a, 2005b; Schwartz \& Martin 2004; Martin 2005),
complex UV emission and absorption lines (e.g., Lequeux et al. 1995;
Kunth et al. 1998; Heckman et al. 1998; Gonzalez Delgado et al. 1998),
extended emission in soft X rays (e.g., Read, Ponman, \&
Strickland 1997; Dahlem, Weaver, \& Heckman 1998; Pietsch et al. 2000;
McDowell et al. 2003; Ehle et al. 2004; Huo et al. 2004; Strickland et
al. 2004a, 2004b; Ehle 2005; Grimes et al. 2005) and at radio
wavelengths (e.g., Hummel, Beck, \& Dettmar 1991; Irwin, English, \&
Sorathia 1999; Irwin, Saikia, \& English 2000; Dahlem et al. 2001) all
confirm this dependence on starburst strength.

A key indicator is 60-to-100 $\mu$m IRAS color,
$S_{60}/S_{100}$, which tracks dust temperature in star-forming
regions. ``Warm'' 60-to-100 $\mu$m IRAS color ---
$S_{60}/S_{100} \ga 0.4$ --- is often used to classify galaxies as
starbursts.  Warm galaxies generally have large IR luminosities ($\ga$
10$^{10.5}$ L$_\odot$),
excesses (L$_{\rm IR}$/L$_{\rm opt}$ $\ga$ 2), 
and galaxy-averaged far-IR (FIR) surface brightnesses (L$_{\rm
  FIR}$/$\pi R_{25}^2$ $\ga$ 2 $\times$ 10$^{40}$ erg~s$^{-1}$
kpc$^{-2}$; e.g., Rossa \& Dettmar 2003a). This is not surprising
because the SFR scales with the IR luminosity according to equation
(5), and the dust temperature is expected to scale with the UV energy
density.
The conditions on the IR luminosity and surface brightness in warm
galaxies translate into SFR $\ga$ 5 M$_\odot$ yr$^{-1}$ and SFR/$\pi
R_{25}^2 > 0.001$ M$_\odot$ yr$^{-1}$ kpc$^{-2}$. Almost all of these
galaxies have extraplanar ionized gas, and most also host a wind. More
than 75\% of ultraluminous IR galaxies (ULIRGs) with IR luminosities
$>10^{12}$ L$_\odot$ ($S_{60}/S_{100} \ga 0.5$) have winds (e.g.,
Rupke et al. 2002, 2005b; Martin 2005). The degree of nucleation of
star formation activity increases with SFR, so it is not surprising to
detect so many winds among ULIRGs (where SFR $\ga$ 100 M$_\odot$
yr$^{-1}$ and $R_* < 1$ kpc).

\subsection{Morphology}

The outflows in most star-forming galaxies have a bipolar distribution
perpendicular to the disk.  Opening angles are $2 \theta \approx
10^\circ$ -- 45$^\circ$ near the base, increasing to $\sim 45^\circ$ --
$100^\circ$ above the disk as expected from simulations (\S 2.4).
But the detailed gas distribution is often complex, shows large
galaxy-to-galaxy variations, and depends on the specific gas phase.
In optical line emission, outflow structures range from the classic
egg-shaped nuclear superbubble of NGC~3079\footnote{NGC~3079 has both
a nuclear starburst and an AGN, but from the morphology and kinematics
of the line emitting gas, Cecil et al. (2001) conclude that this
outflow is starburst, not AGN, powered.}  (Fig.\ 3; Ford et al. 1986;
Veilleux et al. 1994; Cecil et al. 2001) to the bipolar double-loop
morphology of Arp~220 (Heckman, Armus, \& Miley 1987), the biconical
structure of M82 (Fig.\ 4; Bland \& Tully 1988; Shopbell \&
Bland-Hawthorn 1998; Ohyama et al. 2002)
and NGC~1482 (Fig.\ 5$a$; Veilleux \& Rupke 2002), and the frothy and
filamentary morphology of the outflow in the dwarf NGC~1569 (e.g.,
Martin, Kobulnicky, \& Heckman 2002).  The line-emitting structures
are often limb-brightened, indicating that much of the optically
emitting gas resides on the surface of largely hollow structures. In
the few objects examined with the {\em Hubble Space Telescope} (HST)
($\sim$ 0\farcs1), the line-emitting gas is resolved into complexes of
clumps and filaments (e.g., M82) or streams of filamentary strands and
droplets (e.g., NGC 3079) with volume filling factors $f \ga 10^{-3}$.
The morphology and kinematics (discussed in \S 4.4) of this gas
suggest that it originated in the cool disk and became entrained in
the outflow.

\begin{figure} 
\caption{\small NGC 3079 imaged with HST (red for
H$\alpha$+[N~II], green for I-band) and CXO (blue). (a)
large-scale emission across $15\times5$ kpc.  Numerous H$\alpha$
filaments rise above the disk.  Note the V-shaped wind pattern
extending in X rays from nucleus; for clarity, we have suppressed the
diffuse X-ray emission across the superbubble, where it is generally
clumped (see [c]).  (b) The $1\times1.2$ kpc superbubble in
H$\alpha$+[N~II] emission, with log-scaled intensities. It is composed of 4
vertical towers of twisted filaments; the towers have strikingly
similar morphologies.  (c) Close-up of the wind-swept, circumnuclear
region.  Note how X-ray emission (blue) clumps along the optical
filaments of the superbubble at the limit of CXO's
resolution. A prominent dust filament at left drops out of the wind.}
\end{figure}

\begin{figure} 
\caption{\small M82, imaged by the WIYN telescope in
H$\alpha$ (magenta) and HST in BVI continuum colors (courtesy
Smith, Gallagher, \& Westmoquette).  Several of the largest scale
filaments trace all the way back to super-starclusters embedded in the
disk.}
\end{figure}

\begin{figure} 
\caption{\small (a) Starburst galaxy NGC 1482 imaged in H$\alpha$
(red), [N~II] (green), and CXO (blue).  The bar is 3 kpc
long.  For this and the other three panels, N is up and E is left, and
intensities in all bands are log-scaled. (b) Excitation map of
NGC~1482 showing the shock-excited ([N~II] $\lambda$6583/H$\alpha$
$\ga$ 1 in black), biconical structure due to the wind above and below
the star-forming disk ([N~II]/H$\alpha$ $<$ 1 in orange). The scale is
the same as in (a), and the white crosses indicate the locations of
two bright star-forming regions in the disk. (c) Seyfert galaxy
NGC~4388 imaged in H$\alpha$ (red), [N~II] (green), and CXO
(blue).  The bar is 8 kpc long. The kinematics of the extraplanar gas
$\la$ 8 kpc north of the nucleus are dominated by the AGN-driven
outflow. The long H$\alpha$ trail extending further to the north-east
may be due to ram-pressure stripping by the ICM in the Virgo cluster.
(d) Central region of the Circinus galaxy imaged in [O~III] (blue) and
blueshifted (between --150 and 0 km~s$^{-1}$) H$\alpha$ (red).  The
bar is 1 kpc long. The [O~III] emission traces the ionization cone and
conical wind produced by the AGN, whereas the H$\alpha$ emission
east of the nucleus is dominated by the
circumnuclear starburst. Spectacular filaments and bow shocks are seen
on scales of $\sim$ 500 -- 900 pc. }
\end{figure}

\begin{figure} 
\caption{\small Maximum Na~I~D absorption-line outflow velocities as a
function of (a) circular velocities, and (b) star formation rates. Red
skeletal stars are star-forming dwarfs from Schwartz \& Martin (2004)
and red open stars are infrared-selected starbursts from Rupke et
al. (2005a, 2005b).  Filled blue circles and filled black squares are Seyfert
2s and Seyfert 1s from Rupke et al. (2005c). The dashed line in (a)
represents the escape velocity for a singular isothermal sphere with
$r_{max}/r = 10$, whereas the dashed lines in (b) are characteristic
velocities of ram-pressure accelerated clouds (Murray et al. 2005) for
column densities of 10$^{20}$ cm$^{-2}$ (top line) and 10$^{21}$
cm$^{-2}$. }
\end{figure}

These structures vary in vertical extent from $\sim$ 1
kpc to $\ga$ 20 kpc (see Veilleux et al. 2003 and references therein).
Large filaments seemingly unrelated to the nuclear structure sometimes
appear in deeper exposures; e.g., faint X-shaped filaments extend $>8$
kpc from the nucleus of NGC~3079 (Heckman et al. 1990; Veilleux et
al. 1994) and rise $\sim$ 4 kpc from the galactic plane but connect to
the inner ($R \approx 1.5$ kpc) galactic disk rather than to the
nucleus itself.  Cecil et al. (2001) suggest that these filaments have
the shape expected (Schiano 1985) for the contact discontinuity/ISM
associated with lateral stagnation of the wind in the galaxy thick
disk/halo. The lateral extent of the wind is necessarily much smaller
in the disk, where one expects to find a ``ring shock."  This feature
(see \S 2.4) has been detected in a few objects, based on line
emission and warm molecular gas emission (e.g., NGC~253: Sugai,
Davies, \& Ward 2003; see below).

A growing set of $\sim$ 1\arcsec-resolution CXO data show that
the bright, soft X ray and H$\alpha$ filaments in these winds have
strikingly similar patterns on both small and large scales ($\sim$
0.01 -- 10~kpc; e.g., Strickland et al. 2000, 2002, 2004a; Cecil,
Bland-Hawthorn, \& Veilleux 2002a; McDowell et al. 2003).  
This tight optical-line/X-ray match seems to arise from
cool disk gas that has been driven by the wind, with X rays being
emitted from upstream, standoff bow shocks or by conductive cooling at
the cloud/wind interfaces.  This is not always the case for the
fainter soft X-ray emission.  For instance, the X-ray emission near
the X-shaped filaments of NGC~3079 (Fig.\ 3$a$) is not significantly
edge brightened, suggesting a partially filled volume of warm gas
within the shocked wind, not a shell of conductively heated gas
(Cecil, Bland-Hawthorn, \& Veilleux 2002; see also Huo et al. 2004;
Ott, Walter, \& Brinks 2005). In a few objects, absorption by
foreground neutral ISM also affects the distribution of the soft X-ray
emission (e.g., NGC~253: Strickland et al. 2000; M82: Stevens, Read,
\& Bravo-Guerrero 2003).  Intrinsically hard, diffuse X-ray emission
has been detected in a few wind galaxies.  In M82, this gas is more
nucleated but less filamentary than the soft X-ray emission; it
probably traces the hot, high-pressure wind fluid (Griffiths et al.
2000; Stevens et al. 2003; see \S4.9 for more detail).

Several wind galaxies have large radio halos (e.g., M82: $Z \approx 5$
kpc, Seaquist \& Odegard 1991; NGC~253: $\sim$ 9 kpc, Carilli et
al. 1992; NGC~3079: $\sim$ 11 kpc, 
Irwin \& Saikia 2003; NGC~4631: $\sim$ 9 kpc, Hummel \& Dettmar 1990).
Galaxies with thick ionized disks ($S_{60}/S_{100} > 0.4$) also tend
to show extraplanar synchrotron radio emission (e.g., Dahlem et
al. 2001).  Given the significant polarization of the radio emission
and the lack of point-by-point correspondence with the X-ray and
optical line emission, non-thermal synchrotron from magnetized ($B
\approx {\rm few} \times 10 \mu$G), relativistic electrons is the
favored explanation for most of the radio emission.  The pattern of
magnetic field lines in NGC~3079 (Cecil et al. 2001) suggests that the
relativistic electrons are produced in the starbursting disk and then
advected from the disk by the wind.
Some emission may also come from electrons accelerated locally in
internal wind shocks.  Steepening of the spectral index of the radio
continuum emission in M82 (Seaquist \& Odegard 1991), NGC~253 (Carilli
et al. 1992), NGC~4631 (Ekers \& Sancisi 1977; Hummel 1991) and
NGC~891 (Allen, Sancisi, \& Baldwin 1978; Hummel 1991) indicates
energy losses of the electrons on their way from the starburst, either
from synchrotron losses or inverse Compton scattering of the
relativistic electrons against the IR photons produced in the nuclear
region.  In some objects, the relativistic component of the wind
appears to decouple from the thermal component beyond the
H$\alpha$-emitting structures (e.g., NGC~3079, Duric \& Seaquist
1988), perhaps participating in a ``cosmic ray wind'' rather than a
thermal wind (Breitschwerdt \& Schmutzler 1999).

Unambiguous evidence for entrained neutral gas has been detected from
dwarf galaxies (e.g., Puche et al. 1992; Stewart et al. 2000; Schwartz
\& Martin 2004) to ULIRGs (e.g., Heckman et al. 2000; Rupke et
al. 2002, 2005a, 2005b; Martin 2005). This gas often extends up to
several kiloparsec, but morphological constraints are sparse. Detailed
long-slit spectra show some degree of correlation with the warm
ionized gas (\S 4.4). Kinematically disturbed molecular gas has also
been detected in a few GWs.  The best case for molecular gas entrained
in a GW is in M82, where a detailed kinematical decomposition of the
CO gas into wind and disk components reveals a wide-angle (opening
angle $2 \theta \approx 110^\circ$) pattern related loosely to the
outflow seen at other wavelengths (Stark \& Carlson 1984; Seaquist \&
Clark 2001;
Walter, Weiss, \& Scoville 2002). A narrow, shock-excited SiO chimney
extends $\sim$ 500 pc above the disk (Garcia-Burillo et al. 2001), and
is also found in NGC~253 (Garcia-Burillo et al. 2000).  Shocked H$_2$
gas is detected at the base of the outflow of NGC 253 (Sugai et
al. 2003),
consistent with the cloud-crushing model of Cowie, McKee, \& Ostriker
(1981) and Ohyama, Yoshida, \& Takata (2003) where C-type shocks with
$V_{\rm shock} \la $ 40 km~s$^{-1}$ are compressing the star-forming
molecular disk.

GW structures sometimes tilt relative to the minor axis of the host,
and are asymmetric to the nucleus. Tilts and asymmetries near the
starbursts (M82: Shopbell \& Bland-Hawthorn 1998; Galaxy:
Bland-Hawthorn \& Cohen 2003; NGC~253: Sugai et al. 2003) probably
reflect asymmetries in the starbursting population and in the density
distribution of the cool, star-forming disk. Asymmetries on large
scales may be due to density fluctuations in the halo of the host
galaxy, and therefore can be a probe.  The IGM through which the
galaxy is moving may also influence the morphology of the wind
structure on large scales (e.g., radio halo in M82, Seaquist \&
Odegard 1991).

\subsection{Kinematics}

Spectra of the warm ionized component in GWs shows double-peaked
emission-line profiles with a split ranging from a few 10's of km
s$^{-1}$ in some dwarf galaxies (e.g., Marlowe et al. 1995; Martin
1998) to about 1500 km~s$^{-1}$ in NGC~3079 (Filippenko \& Sargent
1992; Veilleux et al. 1994). Constraints on the phase-space
distribution of gas within the outflows are necessary to deproject
observed velocities. The dense 2D spatial coverage of Fabry--Perot and
IF spectrometers is ideal, but usually only
narrow-band imaging and long-slit spectroscopy are available to
constrain the outflow geometry.  The observed kinematic patterns
indicate that most warm gas lies on the surface of expanding
bubbles/ellipsoids or flows along the walls of conical structures. The
substantial broadening often seen in individual kinematic components
indicates that the flow is not purely laminar or that the cone walls
are composites of distinct filaments with a range of velocities.

The most detailed studies of outflow kinematics can be compared
quantitatively with the hydro simulations described in \S 2.4.  The
deprojected outflow velocity in open-ended conical winds often
increases with radius as expected for entrained gas in a free-flowing
wind (e.g., Murray et al. 2005); examples are M82 (Shopbell \&
Bland-Hawthorn 1998) and NGC~3079.  Gas near the top of the partially
ruptured bubble of NGC~3079 is entrained in a mushroom vortex (Cecil
et al. 2001), as predicted theoretically (e.g., Suchkov et al. 1994).
In these open-ended winds, there is a clear correlation between
outflow velocity and gas phase temperature. For example, cool
molecular gas is ejected out of M82 at a maximum deprojected outflow
velocity of $\sim$ 230 km~s$^{-1}$ (Shen \& Lo 1995; Walter et
al. 2002), well below the inferred velocities of the warm ionized gas
(525 -- 655 km~s$^{-1}$; Shopbell \& Bland-Hawthorn 1998). Coronal gas
traced by the O$\:$VI $\lambda$1032 absorption line in NGC~1705 is
more blueshifted than the neutral gas, whereas the warm photoionized
gas appears to have intermediate velocities (Heckman et
al. 2001a). Smaller outflow velocities in the neutral gas relative to
the ionized components are also often seen in LIRGs and ULIRGs (Rupke
et al. 2005b). These systematic kinematic variations with phase
temperature are consistent with entrainment of gas clouds by a hot
wind if the warmer phase has smaller column densities than the cool
gas, perhaps as a result of cloud erosion (\S 2.4).

Entrainment of disk material is supported by other evidence, including
the detection of rotation in the outflow gas (e.g., M82: Shopbell \&
Bland-Hawthorn 1998, Greve 2004; NGC~3079: Veilleux et al. 1994;
NGC~1482: Veilleux \& Rupke 2002) and the field reversal across the
NGC~3079 radio lobe that suggests the return of entrained, magnetized
material to the disk (Cecil et al. 2001).

Wind kinematics are poorly constrained. This fluid, being hot and
tenuous (\S 4.9 and Equation [8]), is hard to detect, so current X-ray
instruments do not constrain its velocity.  A lower limit follows from
the expected terminal velocity of an adiabatic wind at the measured
X-ray temperature $T_X$: $V_W^2 \simeq \alpha~c_s^2 = \alpha~k T_X/\mu
m_p$, where $c_s$ is the isothermal sound speed of the hot phase, $\mu
m_p$, the mean mass per particle, and $\alpha = 2.5 - 5.0$, a scale
factor that depends on the fraction of thermal energy that is radiated
(e.g., Appendix B in Efstathiou 2000). This is a lower bound because
it accounts only for thermal energy and neglects possibly substantial
(e.g., Strickland \& Stevens 2000) bulk motion. Measured X-ray
temperatures in GWs are quite uniform ($\sim$ 0.2 -- 2 $\times$ 10$^7$
K; Martin 1999; Heckman et al. 2000; Strickland et al. 2004a), and
imply $V_W \simeq$ 500 -- 900 km~s$^{-1}$. These are well below
$V_{\infty}$ from
equation (16) unless $\xi/\Lambda \ll 1$.  Note that the X-ray
temperature is weighted by the emission measure, so it probes only
dense shocked material between disk and wind gas. The measured X-ray
temperature is therefore likely a {\em lower} limit to the wind
temperature, and wind velocities are also lower limits.

Our knowledge of the kinematics of neutral gas in GWs has improved
considerably thanks to Na~I~D absorption-line surveys (e.g., Heckman
et al. 2000; Rupke et al. 2002, 2005a, 2005b; Schwartz \& Martin 2004;
Martin 2005).
The profile of the interstellar Na~I absorption feature is fit with
multiple components to determine the bulk and turbulent velocities of
the neutral gas. The resulting distribution of Na~I~D velocities is
skewed to the blue (relative to systemic); this is interpreted as
outflow.  Line full widths at half maximum (FWHMs) average $\sim$ 275 km~s$^{-1}$ in large
starbursts, much larger than the thermal velocity dispersion of warm
neutral gas. This broadening comes from the superposition of distinct
kinematic components with several radial velocities, as seen in the
warm gas phase.  The projected ``maximum'' velocities in the
outflowing components (equal to the centroid velocity plus one-half
the velocity width) average 300 -- 400 km s$^{-1}$, and attain $\sim$
600 km~s$^{-1}$ (although 1100 km~s$^{-1}$ is seen in
F10378+1108). Figure 6$a$ plots the maximum outflow velocities against
host-galaxy mass. A linear least-squares fit suggests that outflow and
circular velocities correlate positively, but this correlation is 
mainly due to the dwarf galaxies (Martin 2005, Rupke et al. 2005b).

Figure 6$b$ compares the outflow velocities with the host galaxy SFR
($L_{\rm IR}$).  There is some indication of a trend of increasing
outflow velocities with increasing SFR, particularly when data from
Schwartz \& Martin (2004) on dwarf galaxies are included. Figure 6$b$
shows that cloud entrainment in a wind (Equation [A5] in Murray et
al. 2005) is able to reproduce these velocities, except perhaps for
F10378+1108 where the large outflow velocity may also require
radiation pressure.
(Martin 2005 has also argued for radiation-pressure driving in other
objects).  UV absorption-line measurements appear to confirm the
positive correlation between outflow velocities and SFR's (Heckman
2004).

\subsection{Mass Outflow Rates \& Energetics}

The multiphase nature of GWs greatly complicates the task of
estimating the outflow masses and energies.  A multiwavelength
approach that considers all phases is essential. Given the
density-squared dependence of the emission measure, diagnostics that
rely on line or continuum emission favor the densest material, yet
this may be only a relatively small fraction of the total mass and
energy.  This is especially relevant for the wind fluid, which is
expected to dominate the energetics of the outflow, but contributes
very little to the emission (Strickland \& Stevens 2000).  Corrections
must be applied for the volume filling factor, $f$.
Constraints on $f$ can be derived from estimates on the volume of the
emitting material and assumptions about its emissivity (temperature).
Measurements that rely on absorbing column densities are less subject
to density inhomogeneities, but require a strong source of background
light and therefore are limited to the brightest parts of the
starburst.  Assumptions about filling factor and outflow extent must
then be made to estimate the mass of outflowing material and the
energetics.

After deprojecting observed velocities, the dynamical timescales of
starburst-driven outflows ($t_{\rm dyn} \approx R/V_{\rm out}$) range
from 0.1 to 10 Myr.  
The mass of warm ionized gas inferred to take part in the outflow is
$\sim$ 10$^5$ -- 10$^6$ M$_\odot$ in dwarf galaxies and $\sim$ 10$^5$
-- 10$^7$ M$_\odot$ in powerful starbursts.  The dynamical timescales
yield mass outflow rates ranging from $\la$ 0.1 M$_\odot$ yr$^{-1}$ to
$\ga$ 10 M$_\odot$ yr$^{-1}$, with a trend for the rate to increase
with increasing SFR. A similar trend may exist between star formation
activity and the amount of extraplanar gas in non-starburst galaxies
(e.g., Dettmar 1992; Rand, Kulkarni, \& Hester 1992;
Miller \& Veilleux 2003a; Rossa \& Dettmar 2003a).

Bulk plus ``turbulent" KEs inferred from optical emission-line spectra
span a broad range: e.g., NGC~1482: $\ga 2 \times 10^{53}$ erg
(Veilleux \& Rupke 2002); NGC~3079: $\sim$ $10^{54}$ erg (Cecil et
al. 2001); M82: $\sim$ $2 \times 10^{55}$ erg (Shopbell \&
Bland-Hawthorn 1998). These are typically several times lower than
thermal energies derived from X-ray data. Luminosities of galactic
X-ray halos scale roughly with IR luminosities or disk SFRs
(Strickland et al. 2004a, 2004b), but the corresponding X-ray cooling
rates amount to $\la$ 10\% of the SN heating rates in these
objects. Cooling by the optical line-emitting material is even
smaller.

Results from Na~I~D studies suggest that GWs entrain considerable
neutral material, $\sim$ 10$^4 - 10^7$ M$_\odot$ (0.001 -- 1.0
M$_\odot$ yr$^{-1}$) in dwarfs and $\sim$ 10$^8 - 10^{10}$ M$_\odot$
($\sim$ 10 -- 1000 M$_\odot$ yr$^{-1}$) in ULIRGs.  These rates
generally exceed the mass injection rate from SNe (Equation [1]).  However,
they are highly uncertain because: (1) the geometry of the neutral
outflow is poorly known (here the simple model of a mass-conserving
free wind from Rupke et al. 2005b is used with the assumption that gas
lies in a thin shell at a radius $\sim$ 5 kpc for the large
starbursts; and at smaller radii for the dwarfs; the estimated masses scale
linearly with this radius); (2) depletion of Na onto grains affects
the strength of the Na~I~D absorption line (here we assume a depletion
factor $\sim$ 9); (3) the ionization potential of Na is low (5.139 eV)
so considerable Na is ionized even when hydrogen is neutral; this must
be accounted for when calculating H$^0$ masses (the ionization
correction is assumed to be close to the Galactic value of $\sim$ 10).

Assuming that these estimates of neutral gas masses are correct, the
ratios of $\dot{M}$ to the global SFRs span $\eta \equiv \dot{M} /
\mathrm{SFR} \approx 0.01 - 10$, consistent with those found by Martin
(1999) for the warm ionized medium of ten galaxies. Parameter $\eta$,
the mass entrainment efficiency, shows no obvious dependence on
SFR except perhaps for a decreasing trend at high SFR ($\ga$ 10 -- 100
M$_\odot$ yr$^{-1}$). The inferred KE
increases with increasing SFR: $\sim$ 10$^{50} - 10^{54}$ erg among
dwarfs but $\sim$ 10$^{56} - 10^{59}$ erg among LIRGs and ULIRGs;
corresponding power outputs are $\sim$ 10$^{36} - 10^{39}$ and
10$^{41} - 10^{44}$ erg~s$^{-1}$, respectively.  Such energies and
powers exceed those of the outflowing warm ionized gas, and imply
thermalization efficiencies $\ga$ 10\%.  Contrary to
some expectations (e.g., Silk 2003), the trend with SFR flattens among
ULIRGs, perhaps due to the complete evacuation of the gas in the
wind's path, a common neutral gas terminal velocity for LIRGs and
ULIRGs, and/or a decrease in the efficiency of thermalization of the
SN energy.

CO studies of GWs are very important because so much mass is required
to see molecular gas that its detection dramatically increases the
inferred energies.  A good illustration is M82: Walter et al. (2002)
deduced from CO observations that $> 3 \times$ 10$^8$ M$_\odot$ of
H$_2$ is involved in the outflow, and its KE $\sim10^{55}$ erg becomes
comparable to the KE in the warm filaments. We discussed in \S3.2 the
GW of the Milky Way, where cold material seems to dominate the KE of
the outflow.  Detailed millimeter studies of a representative set of
GWs will be needed to confirm the dynamical importance of the
molecular gas component.

The latest addition to the mass and energy budgets of GWs is the
coronal ($T \sim 10^5$ K) phase traced by O$\:$VI. Dynamical
information on this component is currently sketchy, outflowing gas
being detected in absorption in only two (dwarf) galaxies so far:
NGC~1705 ($v_{\rm out} \approx 100$ km~s$^{-1}$; Sahu \& Blades 1997;
Heckman \& Leitherer 1997; Heckman et al. 2001) and NGC~625 ($v_{\rm
out} \approx 30$ km~s$^{-1}$; Cannon et al. 2005).
The mass of coronal gas derived from these data is uncertain because
of possible line saturation, ionization corrections, and assumptions
on the gas geometry. In NGC~1569, Heckman et al. (2001a) estimate a
mass of $\sim$ 6 $\times$ 10$^5$ M$_\odot$ and KE of $\sim$ 3 $\times$
10$^{52}$ erg in the coronal phase. These are only $\sim1\%$ of the
values in the warm ionized phase, so the coronal component is
unimportant dynamically. But what about its radiative losses? The O~VI
$\lambda\lambda$1032, 1038 lines are key because they produce $\sim$
30\% of the coronal cooling.
O$\:$VI emission was detected in NGC~4631 (Otte et al. 2003), but not
in NGC~1705 (Heckman et al. 2001a), M82 (Hoopes et al. 2003), and
NGC~891 (Otte et al. 2003).  These measurements limit the radiative
cooling of coronal gas to 10 -- 20\% of the SN heating rate in
these objects. To within a factor of 2, this is identical to the
cooling rate of the X-ray-emitting gas.

\subsection{Escape Fraction}

The fraction of outflowing material that can escape the gravitational
potential of the host galaxy is an important quantity but is difficult
to determine accurately.  It should be recalled that within the
context of CDM theory, the virial radius of the dark halo is
$\sim250h_{70}^{-1}$ kpc for an $L_*$ galaxy, with the IGM beyond.

Can winds reach the IGM? The main uncertainty comes from our lack of
constraints on halo drag. Silich \& Tenorio-Tagle (2001) have argued
that drag may severely restrict a wind and limit its escape fraction.
Drag by a dense halo or tidal debris may be especially important in
high-luminosity starbursts because many are triggered by galaxy
interactions (e.g., Veilleux, Kim, \& Sanders 2002b). Large H~I halos
may also prevent dwarf galaxies from venting a wind (e.g., NGC~4449:
Summers et al.  2003).  Conversely, Strickland et al. (2004b)
suggested that high-latitude hot gas above the disk can actually help
the outflow to escape.

A popular way to estimate the escape fraction is to compare the
outflow velocity with the local escape velocity derived from a
gravitational model of the host galaxy. This is often a simple,
truncated isothermal sphere.  If truncated at
$r_{\rm max}$, then the escape velocity $v_{\rm esc}$ at radius $r$ is
related to the rotation speed $v_c$ and $r_{\rm max}$ by $v_{\rm
esc}(r) = \sqrt{2}~v_c~[1 + ln(r_{\rm max}/r)]^{\frac{1}{2}}$. The
escape velocity is not sensitive to the exact value of $r_{\rm max}/r$
[e.g., for $r_{\rm max}/r = 10 - 100$, $v_{\rm esc} \approx (2.6 -
3.3) \times v_c$]. The curve in Figure 6$a$ is for $r_{\rm max}/r =
10$. If halo drag is tiny, material that exceeds $v_{\rm esc}$
may escape into the IGM. With this
simple assumption, Rupke et al. (2005b) find that $\sim 5 - 10$\% of the
neutral material in starburst-driven winds will escape. 
This may only be a lower limit: much of the gas above $v_{\rm esc}$
may have already mixed with the IGM and would be invisible in Na~I~D
absorption.

Given the correlation between outflow velocity and gas-phase
temperature mentioned in \S 4.4, the escape fraction is surely larger
for warm and hot phases. Indeed, warm gas in several dwarfs (including
possibly M82) exceeds escape (e.g., Martin 1998; Devine \& Bally 1999;
Lehnert, Heckman, \& Weaver 1999).  Similarly, velocities derived in
\S 4.4 from X-ray temperatures, $V_w \approx$ 500 -- 900 km s$^{-1}$,
exceed escape for galaxies with $v_c \approx 130 - 300$ km~s$^{-1}$.
Recall that these are lower limits to the wind terminal velocities,
so galaxies with $v_c \la 130$ km~s$^{-1}$ may not retain hot,
metal-enriched material (Martin 1999). 
As discussed in \S 7.1.2, this galaxy-mass dependence on metal
retention makes definite predictions on the effective yield which
appears to have been confirmed by observations.

\subsection{Energy Injection Zone}
 
HST images of nearby starbursts show that the star-forming regions
can be extremely complex and luminous super-star clusters (SSCs), each
having thousands of young ($\la 50$ Myr) stars within a half-light
radius $\la$ 10 pc.
But the clustered part of star formation accounts for only $\sim$ 20\%
of the integrated UV light in nearby optically-selected starbursts
(Meurer et al. 1995). Most of the star formation is distributed
diffusely in these objects.
It is unclear which mode of star formation (clustered vs.\ diffuse)
drives starburst winds. Both seem to in the outflow from the dwarf
galaxy NGC~1569, whose wind seems to emanate from the entire stellar
disk rather than just the central 100 pc near the SSCs (Martin et
al. 2002). In M82, Shopbell \& Bland-Hawthorn (1998) deduced from the
diameter of the inner outflow cone a relatively large energy injection
zone, $\sim$ 400 pc.  But chimneys at the base of the wind suggest
localized venting of hot gas (Wills et al. 1999; see also Fig.\ 4).
Heckman et al. (1990) used the gas pressure profile (Equation [10]) to
deduce relatively large (a few hundred parsec) vertical sizes for the
energy injection zones in several edge-on galaxies.  But one should be
wary of the possibility that the [S~II] $\lambda\lambda$6716, 6731
emission lines that they used to derive density can be severely
contamined by foreground or background disk emission; Veilleux et al.
(1994) showed that this occurs in NGC~3079.  Additionally, these may
measure only vertical pressure profiles in the foreground/background
galaxy disk, not the pressure profile in the central starburst.
 
Given the dependence of thermalization efficiency on density (\S
2.1.1), in high-density environments diffuse star formation in the
low-density ISM, not clustered star formation, may drive the wind most
efficiently.  Hence, Chevalier \& Fransson (2001) have warned about
using the radio continuum as a proxy for mechanical luminosity in
starburst galaxies. Here, radio emission comes predominantly from
SN remnants that are expanding in the dense ($\ga 10^3 - 10^4$
H atoms cm$^{-3}$) interclump medium of molecular clouds. These
radiate most of their mechanical energy, so they do not drive GWs.  On
the other hand, SNe that detonate in a lower-density medium heat gas
and drive winds, but are largely invisible in the radio.
The large gas densities at the centers of ULIRGs ($\sim$ 10$^{4-5}$
cm$^{-3}$) (Downes \& Solomon 1998; Bryant \& Scoville 1999; Sakamoto
et al. 1999)
seem to imply a large volume filling factor of molecular clouds and
possibly strong radiative losses, perhaps explaining the small mass
entrainment efficiencies in ULIRGs (\S 4.5).
 
\subsection{Excitation Properties \& Evolution}
 
Superbubbles and GWs are time-dependent, dynamic systems.  How an
outflow evolves is tied directly to the history of its energy source,
{\em i.e.} the star formation history of the host (starburst versus
quiescent star formation).  As a GW evolves spatially (\S\S 2.2 --
2.3), the gaseous excitation of any entrained material also changes
because of two processes: (1) photoionization by hot OB stars in the
starburst and by the hot X-ray-emitting wind, and (2) radiative shocks
formed at the interface of the fast-moving wind and the slow-moving
disk ISM.  As expected from radiative shock models
(Dopita \& Sutherland 1995, 1996), the importance of shock-excited
line emission relative to photoionization by the OB stars in the
starburst appears to scale with the velocity of the outflowing gas
(e.g., Lehnert \& Heckman 1996; Veilleux et al. 2003; Rupke et al.
2005b). NGC~3079 is an extreme example of a shock-excited wind nebula,
with outflow velocity of order 1500 km~s$^{-1}$ (e.g., Filippenko \&
Sargent 1992; Veilleux et al. 1994; Cecil et al. 2001).  The
excitation contrast between the star-forming host and the
shock-excited wind material can in principle be exploited to search
efficiently for GWs (Fig. 5$b$; Veilleux \& Rupke 2002).

The dynamical state of an outflow also affects its gaseous excitation:
Compact, pre-blowout superbubbles are less porous to ionizing
radiation from hot stars than fragmented, post-blowout superbubbles or
GWs. Such fine-tuning between outflow velocities and self-shielding of
the starburst may explain why GWs dominated by OB-star photoionization
are rare.  ``Inverted'' ionization cones, where stellar
photoionization dominates over shocks, have been detected in M82. This
wind has cleared channels beyond the two prominent starbursting knots,
allowing ionizing radiation to escape to large radii (Shopbell \&
Bland-Hawthorn 1998).

\subsection{Wind Fluid}

There is little direct evidence for the wind fluid in starbursts,
because it is tenuous and hot, and therefore a poor X-ray radiator
(e.g., Strickland \& Stevens 2000; Strickland et al. 2000).  The best
evidence for it is in M82 (Griffiths et al. 2000; Stevens et
al. 2003), where the hottest gas is $\la$ 75 pc from the center, and
has $T \sim 4 \times 10^7$ K and pressure $P/k \simeq 10^9$ K
cm$^{-3}$ if the X rays are mostly thermal (Griffiths et
al. 2000). Then the hot fluid is overpressured relative to the disk
ISM and drives the large-scale wind.

Hard (1 -- 6 keV) X-ray emission is resolved in the dwarf galaxy
NGC~1569 (Martin et al. 2002) and its temperature exceeds escape
velocity ($\sim$ 80 -- 110 km~s$^{-1}$; Martin 1999). Interestingly,
the spectral softening of X rays with radius that is expected from
adiabatically cooling winds (e.g., Chevalier \& Clegg 1985) is not
seen, perhaps because of large mass loading.

The metal content of the X-ray-emitting gas may constrain mass
loading, although significant theoretical and observational
uncertainties remain.  As discussed in \S 2.1.1, SN explosions are
expected to dominate when GWs develop. So, the metallicity of the wind
fluid is regulated by SNe yields. Unfortunately, the oxygen and iron
yields of massive stars are only known to an accuracy of $\sim2 - 3$
because of uncertainties in the critical $^{12}$C($\alpha,
\gamma$)$^{16}$O reaction rate, and on the mass limit above which
stars do not contribute to the yield (considerable reimplosion of
heavy elements may affect stars of $\ga$ 30 M$_\odot$; e.g., Woosley
\& Weaver 1995).
Determining the metallicity of X-ray-emitting gas observationally is
notoriously difficult because of uncertainties in the atomic physics
and because of the degeneracies inherent in fitting multi-component
spectral models to data of low spectral resolution. The X-ray-emitting
gas is a multi-phase medium with a range of temperatures, densities,
and absolute/relative metal abundances, possibly located behind cool
absorbing material of unknown metallicity and column density. The
problem is therefore under-constrained and one must assume many
unknowns.
Presently there seems to be evidence that the $\alpha$/Fe ratio is
slightly super-solar in the inner wind of M82 (Stevens et al. 2003;
Strickland et al. 2004a) and in the wind filaments of NGC~1569 (Martin
et al. 2002), as expected if stellar ejecta from SNe II contribute to
the wind fluid.  If confirmed, these modest $\alpha$/Fe enrichments
would further support the idea that mass loading by disk material
contributes significantly to the X-ray emission (recall the kinematic
evidence for disk mass loading in \S4.4).
Martin et al. (2002) compared these measurements with predictions from
SNe models of Woosley \& Weaver (1995) to estimate a mass-loading
factor of $\sim$ 10 in NGC~1569. 

\subsection{Dust}

Evidence is mounting that dust is often entrained in GWs (dust in the
wind of our Galaxy was discussed in \S 3.2). Far-IR maps of a few GWs
show extended cold dust emission along the galaxy minor axis,
suggesting entrainment (e.g., Hughes, Robson, \& Gear 1990; Hughes,
Gear, \& Robson 1994; Alton, Davies, \& Bianchi 1999; Radovich,
Kahanp\"a\"a, \& Lemke 2001).
Color maps reveal elevated dust filaments in several GWs (e.g.,
NGC~1808: Phillips 1993; M82: Ichikawa et al. 1994; NGC~253: Sofue,
Wakamatsu, \& Malin 1994; NGC~3079: Cecil et al. 2001).
In a few systems, including M82, extended polarized emission along the
outflow axis indicates dust (e.g., Schmidt, Angel, \& Cromwell 1976;
Scarrott et al. 1991, 1993; Alton et al. 1994; Draper et al. 1995).
Extended red emission, a broad emission band commonly seen in Galactic
reflection nebulae, exists in the halo of M82 (Perrin, Darbon, \&
Sivan 1995; Gordon, Witt, \& Friedmann 1998).  Far-UV maps of M82 made
with the {\em Ultraviolet Imaging Telescope} reveal a UV-bright
southern cone that is consistent with scattering by dust in the wind
(e.g., Marcum et al. 2001); recent {\em Galaxy Evolution Explorer}
(GALEX) data confirm this conclusion (Hoopes et al. 2005). Other
support for dusty outflows comes from the strong correlation between
nuclear color excesses, $E(B - V)$, and the equivalent widths of
blueshifted low-ionization lines in star-forming galaxies at low
redshifts (e.g., Armus, Heckman, \& Miley 1989; Veilleux et al. 1995b;
Heckman et al.  2000).

Optical measurements can estimate dust masses if one knows the
geometry of the dust filaments and the amount of foreground starlight
and forward scattering. The dust mass in an outflow can best be
estimated from far-IR data, despite uncertainties associated with the
disk/halo decomposition and assumptions about the dust temperature
distribution and emissivity law.
Alton et al. (1999) estimate that $\sim$ 2 -- 10 $\times$ 10$^6$
M$_\odot$ of dust is outflowing from M82. Radovich et al. (2001) used
the same technique to derive 0.5 -- 3 $\times$ 10$^6$ M$_\odot$ of
outflow into the halo of NGC~253.  The dynamical times yield dust
outflow rates of $\sim$ 1 M$_\odot$ yr$^{-1}$.

The fate of this dust is uncertain; there are no direct constraints on
its kinematics in outflows. However, the short sputtering timescale
for silicate/graphite dust in GWs (e.g., Popescu et al. 2000; see also
Aguirre 1999) suggests that grains of diameter $\la 0.3 \mu$m would
not survive long if in direct contact with the wind.  Moreover, there
does not seem to be a tight spatial correlation between the
extraplanar warm ionized medium of starbursts and quiescent galaxies
and the dust filaments (e.g., Cecil et al. 2001; Rossa et
al. 2004). Most likely, the dust is embedded in the neutral or
molecular component of the outflow and shares its kinematics. Assuming
a Galactic gas-to-dust ratio, the neutral gas outflow rates in LIRGs
and ULIRGs (Rupke et al. 2005b) translate into dust outflow rates of
$\sim$ 0.1 -- 10 M$_\odot$ yr$^{-1}$. Given the kinematics of the
neutral gas and assuming no halo drag, $\sim 5-10$\% of the entrained
dust may escape the host galaxy.  Wind-driven ejection of dust from
galaxies may feed the reservoir of intergalactic dust (e.g., Stickel
et al. 1998, 2002), although tidal interactions and ram-pressure
stripping are also efficient conveyors of dust into the ICM of rich
clusters.

\section{LOCAL CENSUS OF LARGE-SCALE OUTFLOWS IN ACTIVE GALAXIES}

The same methods detect outflows in nearby active galaxies and
starburst-driven winds, so the same selection effects and
observational limitations discussed in \S4.1 apply. However, contrary
to starburst-driven winds, winds associated with AGN need not be
perpendicular to the galactic disk. Outflows directed close to the
galactic disk are likely to be made very luminous by the high ambient
densities. They are more easily observed in near face-on galaxies
(e.g., M51: Cecil 1988; NGC~1068: Cecil, Bland, \& Tully 1990).  In a
few systems, the orientation of an inclined AGN disk can be determined
independently from maser spots.

\subsection{Detection Rate}

As we discuss below, there is ample evidence for outflows in AGN from
sub-kpc to galactic scale.  But one must be aware of processes that
may bias the interpretation of bipolar emission.  Several active
galaxies display kiloparsec-scale ``ionization cones''
that align with the radio axes of AGN rather than with the principle
axes of the host galaxy (e.g., Wilson \& Tsvetanov 1994; Kinney et
al. 2000).  They arise when ISM is illuminated by hard radiation from
the AGN. Gas within the cones is more highly ionized than outside.
Because the gas kinematics are unaffected by ionization, one should
search for kinematic signatures to evaluate the frequency of
occurrence of AGN outflows.  Often, both outflows and ionization cones
are present, e.g. NGC~1068 (Pogge 1988; Cecil et al. 1990), NGC~1365
(Veilleux et al. 2003), and NGC~4388 (Veilleux et al. 1999),
emphasizing the need for high-quality kinematic data.

Circumnuclear starbursts in active galaxies are another complication.
Half of all nearby, optically-selected Seyfert 2 galaxies also host a
nuclear starburst (see, e.g., Cid Fernandes et al. 2001 and Veilleux
2001 and references therein); the fraction is larger in IR-selected
systems.  The sustaining conditions for nuclear activity (e.g., deep
central potential with a reservoir of gas) also favor starburst
activity and perhaps trigger starburst-driven GWs.  This symbiotic
relation between starbursts and AGN therefore complicates
interpretation of GWs in AGN/starburst composites.  For example,
bipolar ionization cones can also arise in pure starburst galaxies
(M82: Fig.\ 15 of Shopbell \& Bland-Hawthorn 1998).

With these caveats, we revisit the detection of winds in active
galaxies.  Zensus (1997) reviewed the evidence for outflows in
powerful radio-loud systems.  Here we focus on Seyfert galaxies.
Linear radio features suggestive of jet-like ejections on sub-kpc
scales have long been known in 20 -- 35\% of Seyferts (e.g., Ulvestad
\& Wilson 1989; Morganti et al. 1999 and references therein); this is
a lower limit because of limitations in spatial resolution ($\sim$
1\arcsec), sensitivity, and de-projection.  New high-resolution images
with the {\em Very Large Baseline Array} (VLBA) have indeed revealed
jet-like outflows in previously unresolved sources and in
low-luminosity AGN (Ho \& Ulvestad 2001).  As we describe below,
optical studies of several Seyferts with linear radio structure show
signs of jet-ISM interaction on a scale of tens of pc. Measurements of
proper motion in a few bright sources confirm outwardly moving radio
knots (e.g., Middelberg et al.  2004 and references therein). A jet
interpretation is sometimes favored even when the radio emission is
unresolved at the VLBI scale (e.g., Anderson, Ulvestad, \& Ho 2004).
Low-power, sub-kpc jets may exist in most Seyfert galaxies.  Any
thermal wind component cannot be established from the radio
observations, yet there are speculations that it can be large
(Bicknell et al. 1998)(Section 5.3).

Many active galaxies also show signs of large-scale, galactic
outflows.  In an optical study of a distance-limited sample of 22
edge-on Seyfert galaxies, Colbert et al. (1996a) found that $>$ 25\%
have kinematic signatures of outflows or extraplanar line emission out
to $\ga$ 1 kpc. The existence of extraplanar radio emission in 60\%
(6/10) supports this claim (Colbert et al. 1996b). Morphology and
orientation suggest that this emission comes mainly from AGN-driven
outflows (see \S 5.2).  This detection rate is a lower limit because
of sensitivity and selection effects. For instance, Baum et al. (1993)
detected kiloparsec-scale extranuclear radio emission in $\sim$ 90\% (12/13)
of their objects, a sample of slightly more powerful, face-on
Seyferts.  The difference in detection rate may be due to small number
statistics, differences in sample selection, or may indicate that some
of this emission is associated with starburst-driven
winds. Contamination by starburst-driven winds may also explain the
large fraction (11/12 $\approx$ 90\%) of starburst/Seyfert 2
composites with extended X-ray emission (Levenson, Weaver, \& Heckman
2001a, 2001b).
There is now irrefutable evidence that powerful AGN-driven outflows
sometimes coexist with starburst-driven winds (e.g., NGC~3079: Cecil
et al. 2001).

\subsection{Morphology}

Contrary to what we know of starburst-driven winds, AGN outflows are
oriented randomly relative to the major axis of the host galaxy (e.g.,
Ulvestad \& Wilson 1984; Kinney et al. 2000).
Several well-known active galaxies, including NGC~1068 (Cecil et
al. 1990), harbor a wide-angle outflow whose axis is not perpendicular
to the galaxy disks and therefore interacts strongly with the disk
ISM. Outflows near the disk plane are often well collimated (e.g., NGC
4258: Cecil et al 2000; ESO 428-G14: Falcke, Wilson, \& Simpson 1998),
until they collide with dense disk clouds or until they rise into the
halo (e.g., NGC~4258: Wilson, Yang, \& Cecil 2001). When the jet is
drilling through the disk, detailed correspondence between radio and
optical emission-line features indicates strong jet/ISM interactions
over tens of pc (e.g.,
Falcke et al. 1998; Schmitt et al. 2003a, 2003b and references
therein).  The radio jets compress and shock ambient gas, enhancing
line emission that may dominate the morphology of the NLR. The
well-known correlation between radio and NLR luminosities supports a
dynamical connection between the two (e.g., de Bruyn \& Wilson 1978;
Wilson \& Willis 1980).
A few jet deflections by ISM clouds are also seen (e.g., NGC~1068:
Gallimore, Baum, \& O'Dea 1996b; NGC~4258: Cecil et al. 2000;
NGC~4151: Mundell et al. 2003). Jet/ISM interaction has also been
mapped in X rays thanks to
CXO (e.g., Young, Wilson, \& Shopbell 2001; Wilson et
al. 2001; Yang, Wilson, \& Ferruit 2001).

In many edge-on systems, the radio structure has a linear
or elongated morphology on subkiloparsec scale, but beyond the disk becomes
more diffuse and wide-angled. The change may arise from the vertical
pressure gradient in the surrounding ISM or from momentum loss within
the sub-kpc NLR.  The morphologies of the warm and hot ionized
extraplanar gas are often correlated. They are distributed in a broad
cone near the base of the outflow (e.g., NGC~2992: Colbert et al.
1998; Allen et al. 1999; Veilleux, Shopbell, \& Miller 2001; Circinus:
Veilleux \& Bland-Hawthorn 1997; Smith \& Wilson 2001) but become
filamentary above the disk (e.g., NGC~4388: Veilleux et al. 1999,
Fig.\ 5$c$; NGC~5506: Wilson, Baldwin, \& Ulvestad 1985). Spectacular
bow shocks and finger-like structures are seen optically in the
Circinus galaxy, perhaps due to the abnormally high gas content of
this object (Fig.\ 5$d$; Veilleux \& Bland-Hawthorn 1997). The
correspondence between the extraplanar radio plasma and optical
line-emitting material is often not as tight as that seen in the disk.
Outflows with optical conical geometries may have very different radio
morphologies: e.g., edge-brightened radio bubbles in NGC~2992 (Wehrle
\& Morris 1988) and Circinus (Elmouttie et al. 1998) but lumpy and
filamentary radio structures in NGC~4388 and NGC~5506 (Colbert et al.
1996b). The extraplanar emission at radio wavelengths (where the
foreground disk barely influences the emission) is sometimes lop-sided
(e.g., NGC~4388) from asymmetric energy injection at the source or
from asymmetric ISM on small scales.

\subsection{Kinematics}

Spatially resolved outflows are found on all scales in AGN.
Relativistic outflows on parsec scales in powerful radio-loud sources
are well established (Zensus 1997; Worrall \& Birkinshaw 2004 and
references therein).  Recently, proper motions of radio components
have been measured with VLBI in a few Seyfert galaxies (e.g.,
Middelberg et al. 2004 and references therein). These studies show
that the outward motion of the radio components in these objects is
non-relativistic ($\la 0.25c$) on pc scales. There is now unambiguous
kinematic evidence that these radio jets transfer momentum and energy
to the ambient gas and drive some of the large-scale outflows seen in
radio-quiet and radio-loud objects.

Gas in the NLR ($\sim$ 10~pc $-$ 1~kpc) and extended NLR (ENLR; $\ga$
1 kpc) is an excellent tracer of this jet/ISM interaction.  The good
match between nuclear emission-line widths and bulge gravitational
velocities suggests that the gravitational component dominates in most
Seyferts (Whittle 1985; Wilson \& Heckman 1985; Veilleux 1991; Whittle
1992b; Nelson \& Whittle 1996). But Seyferts with linear radio
structures have long been known to have emission lines with complex
profiles (e.g., Veilleux 1991) and supervirial velocity widths (e.g.,
Whittle 1992a) that implicate an additional source of KE in the NLR.
Detailed long-slit spectra from the ground and
from HST have presented evidence for a dynamical connection between
the NLR and radio components in many of these galaxies (e.g.,
Whittle \& Wilson 2004 and references therein). The complete
spatio-kinematic coverage afforded by Fabry--Perot and integral-field
spectrometers has constrained efficiently the intrinsic, 3D velocity
fields of the outflowing ionized gas
(e.g., Ferruit et al. 2004; Veilleux et al. 2002a and references
therein).  The signatures of jet-driven kinematics seen in some
Seyfert galaxies are also detected in several powerful radio sources,
particularly in compact steep-spectrum radio galaxies and quasars
(e.g.,
Baum, Heckman, \& van Breugel 1992;
Gelderman \& Whittle 1994; McCarthy, Baum, \& Spinrad 1996;
Best, R\"ottgering, \& Longair 2000; Sol\'orzano-I\~narrea, Tadhunter,
\& Axon 2001; O'Dea et al. 2002).

This large data set indicates that expanding radio lobes (Pedlar,
Dyson, \& Unger 1985) or bow shocks/cocoons driven by radio jets
(Taylor, Dyson, \& Axon 1992; Ferruit et al. 1997; Steffen et
al. 1997a, 1997b) accelerate some of the line-emitting gas to $\sim$
100 -- 1000 km~s$^{-1}$.  The fate of the gas clouds --- whether
undisturbed, destroyed, or accelerated --- depends on factors such as
the cloud mass, jet energy flux, and interaction geometry. Dense,
molecular clouds in Seyfert galaxies can deflect radio jets by a
significant angle without experiencing significant damage (e.g.,
NGC~1068: Gallimore et al. 1996a, 1996b; NGC~4258: Cecil et al. 2000).
Jet-cloud interactions can be used to deduce key properties of the
jet. In their analysis of the jet-molecular cloud interaction in the
NLR of NGC~1068, Bicknell et al. (1998) argued that this jet (and
perhaps those in other Seyferts) is heavily loaded with thermal gas
and has low bulk velocities ($\sim$ $0.06c$), contrary to jets in
powerful radio galaxies. Not surprisingly, the jets in Seyfert
galaxies often deposit an important fraction of their KE well within
$\sim$ 1 kpc of the nucleus. The radial velocities of the
emission-line knots in the NLRs of Seyfert galaxies show a slight
tendency to increase out to $\sim$ 100 pc from the nucleus then
decrease beyond, whereas the line widths of the knots decrease
monotonically with increasing distance from the nucleus (e.g.,
Crenshaw \& Kraemer 2000; Ruiz et al. 2005). Deceleration beyond
$\sim$ 100 pc is likely due to drag from ambient gas.

Ram pressure from radio jets/lobes may not always dominate the
acceleration of line-emitting gas in Seyfert galaxies. High-resolution
studies with HST sometimes fail to find a one-to-one correspondence
between the NLR cloud kinematics and the positions of the radio knots.
Ram pressure from a diffuse, highly ionized wind or radiation pressure
by the AGN radiation field has been suggested as the possible culprit
in these cases (Kaiser et al. 2000; Ruiz et al. 2001). The
high-velocity ($\sim$ 3000 km~s$^{-1}$) line-emitting knots detected
in NGC~1068 (Cecil et al. 2002b; Groves et al. 2004) may be explained
by radiation pressure acting on dust grains in the clouds (Dopita et
al. 2002). These knots may correspond to the well-known absorbers seen
projected on the UV continua of some AGN (Crenshaw et al. 2003).
Bright nuclear emission-line knots detected in several other nearby
Seyfert galaxies may also contribute to the population of intrinsic UV
absorbers (Crenshaw \& Kraemer 2005; Ruiz et al. 2005).

Wind ram pressure or radiation pressure may also be responsible for
blueshifted ($\sim$ 100 -- 1000 km~s$^{-1}$) neutral material detected
in several AGN-dominated ULIRGs (Rupke, Veilleux, \& Sanders 2005c), because few of
them show jet-like radio structures.  There is some evidence for
higher average and maximum velocities in Seyfert-2 ULIRGs than in
starburst-dominated ULIRGs, although the evidence for a strong
influence of the AGN on these outflows is inconclusive.  The situation
is quite different among Seyfert-1 ULIRGs, where the outflows are
driven mostly or solely by the AGN (Rupke, Veilleux, \& Sanders 2005c;
Fig. 6$a$). Similarly, nuclear activity is almost certainly
responsible for the broad, blueshifted H~I absorption wings detected
in a growing number of compact radio galaxies. But here, jet-driven
acceleration is implied (e.g., Morganti et al. 2003).

In Seyferts where the energy injection rate of the jet or wind
suffices to eject radio/thermal plasma from the host galaxy, warm
line-emitting gas is entrained into wide-angle outflows. The
intrinsic, 3D velocity field of the line-emitting gas over $\ga1$ kpc
indicates roughly conical radial outflow with opening angles $2 \theta
\approx 60^\circ - 135^\circ$ and hollow (NGC~1365: Hjelm \& Lindblad
1996; NGC~2992: Veilleux et al. 2001), partially filled (NGC~5506:
Wilson et al. 1985), or filamentary geometry (Circinus: Veilleux \&
Bland-Hawthorn 1997; NGC~4388: Veilleux et al. 1999). Deprojected
velocities are $\sim$ 100 -- 500 km s$^{-1}$ in Seyferts, sometimes
larger in radio galaxies. The geometry of the kinematic structures on
large scales depends not only on the energy source but also on the
galactic and intergalactic environment (e.g., the abnormally large gas
content of Circinus may explain the peculiar filamentary morphology of
its outflow).  Rotation is sometimes detected in the velocity field of
the outflowing gas (e.g., Circinus: Veilleux \& Bland-Hawthorn 1997;
NGC~2992: Veilleux et al. 2001), confirming that most of the
line-emitting material comes from the disk ISM.

\subsection{Mass Outflow Rates \& Energetics}
    
Kinematic deprojection is an essential prerequisite to meaningful
dynamical analysis of AGN-driven winds.  Data with complete
spatio-kinematic coverage should be used to derive their masses and
KE. We therefore discuss nearby ($z \la 0.1$) objects for which
spatially resolved kinematic data are available. The broad range of
scales and velocities discussed in \S\S 5.2 and 5.3 implies dynamical
times for the entrained line-emitting material of $\sim$ 10$^4$ --
10$^6$ years in the NLR (including the 3000 km~s$^{-1}$ knots in
NGC~1068) and $\sim$ 10$^6$ -- 10$^7$ years in the ENLR.

Deriving ionized masses for AGN outflows follows the same steps as for
starburst-driven winds, and therefore again favors high-density
material (\S4.5).
Given uncertain volume filling factors, the warm ionized gas masses
derived from optical data on Seyferts, $\sim$ 10$^5$ -- 10$^7$
M$_\odot$, are probably accurate to no better than about $\pm 0.5$
dex. The corresponding mass outflow rates, $M/t_{\rm dyn} \approx 0.1
- 10~M_\odot$~yr$^{-1}$, generally exceed tenfold the mass accretion
rates necessary to fuel the AGN, indicating strong mass loading of the
outflow by the galaxy ISM.

The KE of the warm ionized component in Seyfert galaxies is $\sim$
10$^{53}$ -- 10$^{56}$ erg, including both bulk and ``turbulent''
(spatially unresolved) motions. This range is similar to that in
starburst-driven winds. Dynamical timescales yield KE outflow rates of
$E_{\rm kin}/t_{\rm dyn}$ $\approx$ 10$^{40} - 10^{43}$ erg~s$^{-1}$.
The power radiated by Seyferts in line emission is 10 times larger
and $\sim$ 10$^2 - 10^4$ times larger than in the radio (e.g., Fig.~1
of Wilson, Ward, \& Haniff 1988).  Simple jet-driven models (Blandford
\& K\"onigl 1979a, 1979b; Wilson 1981) reproduce these derived values
using reasonable efficiency factors to relate jet and radio powers.

If the efficiency of energy/momentum transfer to the ambient material
is nearly constant, both the entrained mass and KE scale with AGN
luminosity (e.g., Baum \& McCarthy 2000).  Powerful quasars and radio
galaxies with $L_{\rm bol} \sim 10^{45} - 10^{47}$ erg~s$^{-1}$ (not
discussed so far) seem to have KE outflow rates that exceed those of
Seyferts by several orders of magnitude.  The KE of some outflows is
comparable to the gravitational binding energy of gas in the host
($10^{58} - 10^{60}$ erg). Comparable energy is stored in the large
radio lobes of radio-loud objects: $\gamma/(\gamma - 1) P V \sim
10^{58} - 10^{61}$ erg, where $P$ and $V$ are the pressure and volume
of the radio lobes, respectively, and $\gamma = 4/3$ is the mean
adiabatic index of the relativistic fluid.
Much of the energy dissipated by the AGN accretion disk appears to
find its way into jet mechanical luminosity: powerful radio sources
are Eddington-tuned engines with jet powers $Q \sim 0.1~L_{\rm
Edd}$ (Willott et al. 1999; see also
Falcke, Malkan, \& Biermann 1995).

Outflowing neutral gas detected in absorption in H~I and Na~I~D in
many AGN may add significant energy.  Unfortunately, the mass of
entrained neutral gas in these objects is poorly constrained because
its location is unknown. The mass-conserving wind model of Rupke et
al. (2005b; $R_{\rm outer} = 5$ kpc) applied to the Na~I~D absorption
data on Seyfert 2 ULIRGs yields neutral mass outflow rates similar to
those in starburst-driven winds, {\em i.e.} $\sim$ 10 -- 1000
M$_\odot$ yr$^{-1}$.
With this assumption, neutral-gas KEs and rates of $\sim$ 10$^{56}$ --
10$^{60}$ erg and $E_{\rm kin}/t_{\rm dyn}$ $\approx$ $10^{41} -
10^{45}$ erg~s$^{-1}$ are derived for Seyfert-2 ULIRGs, suggesting
that this material may be a very important dynamical component of
AGN-driven outflows.

\subsection{Energy Injection Zone}

An upper limit to the size of the energy injection zone of AGN-driven
winds can be derived from the geometry of the optical outflow.
Injection zones are often small enough to exclude a starburst origin
for the wind. A good example is the Circinus galaxy (Fig. 5$d$), where
the base of the outflow cone extends over $\la$ 20 pc, 10\% of the
size of the starburst (e.g.,
Smith \& Wilson 2001 and references therein).  The same technique has
been used to measure the size of the primary energy source for the
outflows in NGC~1068 ($\la$ 50 pc; e.g., Cecil et al. 1990; Arribas,
Mediavilla, \& Garcia-Lorenzo 1996; Garcia-Lorenzo, Mediavilla, \&
Arribas 1999), NGC~2992 ($\la$ 100 pc; Veilleux et al. 2001;
Garcia-Lorenzo, Arribas, \& Mediavilla 2001), and in several other
jet-induced outflows (e.g., Whittle \& Wilson 2004).

More accurate assessment of the energy source in AGN-driven winds
often relies on detailed VLA, MERLIN, and VLBA radio maps of the
central regions of AGN, down to scales below tens of parsecs.  Radio
emission generally resolves into a few compact sources, occasionally
accompanied by elongated, jet-like features.  On VLA scales, approximately
half of all Seyfert galaxies contain at least one component with a
flat or inverted radio spectrum ({\em i.e.} spectral slope $\alpha \ge
-0.20$; e.g., Ulvestad \& Ho 2001).  Radio sizes are uncorrelated with
Seyfert type, as expected from the unified scheme of AGN. On VLBI
scales, $\ga80$\% of sources are flat-spectrum (Middelberg et
al. 2004; Anderson, Ulvestad, \& Ho 2004), are often unresolved or
only slightly resolved ($\sim$ 1 pc), and have brightness temperatures
$>10^7$ K. Synchrotron self-absorption from the base of the jet is the
usual explanation for compact and flat- or inverted-spectrum radio
emission from the central engines of radio galaxies. The same scenario
is likely in some Seyfert galaxies and low-luminosity AGN.  Free-free
absorption by the nuclear torus or NLR may also occur. In NGC~1068,
the flat-spectrum source has modest brightness temperature ($\sim$
10$^6$ K) and extends over $\sim$ 1 pc perpendicular to the jet axis;
this emission is believed to be optically thin, free-free emission
from the ionized inner edge of the torus (Gallimore, Baum, \& O'Dea
1997).  Regardless of the exact origin of the emission, the compact
flat-spectrum component is thought to be the main energy source of the
outflow, because of frequent alignment of the various radio components
and from direct measurements of proper motions in a few objects
(Middelberg et al. 2004 and references therein; see \S 5.3 above).

 \subsection{Source of Ionization}
 
It is now generally agreed that the gas in Seyferts is photoionized by
the nucleus, although whether spatial variations in line ratios result
from a ``classical'' range of ionization parameters (e.g. Davidson \&
Netzer 1979), a range in the numbers of illuminated ionization- and
matter-bounded clouds (Binette, Wilson, \& Storchi-Bergmann 1996), or
a dusty plasma at high ionization parameter (Dopita et al. 2002) are
still discussed. The detection of X-ray cones coincident with the
emission-line ionization cones in several Seyferts rules out
photoionizing shocks as the primary source of ionization (e.g.,
Kinkhabwala et al. 2002; Iwasawa et al. 2003; Yang et al. 2001),
although the relative importance of shocks almost certainly increases
with jet power and the degree of interaction with the ambient medium
(e.g., compact radio galaxies; Tadhunter 2002).

\subsection{Dust}

Dust is present in AGN outflows.  Its reddening is evident in the
narrow-line spectrum of radio-quiet and radio-loud galaxies (e.g.,
Osterbrock 1989 and references therein)
and in the broad absorption line clouds of BAL QSOs (e.g., Crenshaw et
al. 2003 and references therein). Dust has often been invoked to
explain the extended blue wings in the narrow-line profiles (e.g,
Veilleux 1991 and references therein), and in the IR continuum
emission of several AGN (Barvainis 1987; Sanders et al. 1989; Pier \&
Krolik 1992).  But more relevant to outflowing dust is that the
correlation between color excess, $E(B - V)$, and the strength of the
Na~I~D line noted for starburst galaxies is even stronger for AGN
(Veilleux et al. 1995b).  Dust outflow rates of $\sim$ 0.1 -- 10
M$_\odot$ yr$^{-1}$ are inferred from the neutral-gas mass outflow
rates discussed in \S 5.4, assuming the Galactic gas-to-dust
ratio. These are comparable to the dust outflow rates in starburst
galaxies (\S 4.10).

\section{WINDS IN THE DISTANT UNIVERSE}
    
\subsection{High-Redshift Galaxies}

There is evidence for winds in the spectra of several $z >$ 1
galaxies.  Low-ionization interstellar absorption lines that are
blueshifted by hundreds of km~s$^{-1}$ relative to systemic
velocities, and Ly$\alpha$ emission lines similarly shifted redward,
have been detected in most $z \sim 3 - 4$ Lyman break galaxies (LBGs;
e.g., Lowenthal et al. 1997;
Pettini et al. 2000, 2001, 2002; Adelberger et al. 2003; Shapley et
al. 2003), in several gravitationally lensed Ly$\alpha$-emitting
galaxies at $z \sim 4 - 5$ (e.g., Franx et al. 1997; Frye, Broadhurst,
\& Benitez 2002), and in many luminous IR galaxies at $z \ga 2$ (e.g.,
Smail et al. 2003; Swinbank et al. 2005). Ly$\alpha$ emission with red
asymmetric or P Cygni-type profiles is also commonly seen in $z \ga 5$
Ly$\alpha$-emitting galaxies (e.g., Dey et al. 1998; Ellis et
al. 2001;
Dawson et al. 2002; Ajiki et al. 2002).  A few isolated $z > 1$
line-emitting ``blobs'' and extended circumnuclear nebulae with
complex structured velocity fields have been tentatively interpreted
as spatially resolved large-scale winds driven by starbursts or AGNs
(e.g., Francis et al. 2001;
Ohyama, Taniguchi, \& Shioya 2004 and references therein). 

Prevalent winds in high-$z$ galaxies, particularly in LBGs, are
to be expected given their large surface densities of star formation:
the typical size of the star-forming cores in LBGs is $r_{\rm half}
\sim$ 1.6 $h_{70}^{-1}$ kpc (e.g., Giavalisco et al. 1996)
and their SFRs are $\sim$ 1 -- 100 M$_\odot$ yr$^{-1}$. Thus, they
have star formation surface densities $>1~M_\odot$ yr$^{-1}$
kpc$^{-2}$. However, one should be cautious when applying local wind
criteria to high-$z$ galaxies because there are significant structural
differences in starbursts between the two epochs.  The energy
injection zone in local starbursts is much smaller than the galaxy
half-light radius and generally centered on the nucleus, whereas the
starbursts in LBGs appear to be galaxy wide.  The prominent disks of
local starbursts were probably not yet assembled in $z \sim 3$ LBGs,
perhaps resulting in more spherical winds than in local starbursts.

A well-studied wind in the gravitationally lensed LBG MS~1512$-$cB58
(Pettini et al. 1998, 2000, 2002) has bulk outflow velocity $\sim$ 255
km~s$^{-1}$ on the basis of the positions of the low-ionization absorption
lines relative to the rest-frame optical emission lines.
Pettini et al. (2002) note that its outflow kinematics are remarkably
symmetric: The profiles of the absorption lines from the different
ionization stages are broadly similar, spanning a range of velocities
of $\sim$ 1000 km~s$^{-1}$, whereas the receding portion of the outflow
as mapped by the redshifted backscattered Ly$\alpha$ emission also has
essentially the same kinematics.  The derived mass outflow rate
($\sim$ 70 M$_\odot$ yr$^{-1}$) exceeds the SFR of this galaxy (SFR
$\approx 40$ M$_\odot$ yr$^{-1}$), so 
this outflow may have had a strong impact on the chemical evolution of
the host galaxy. The entrained wind material appears to have enough
energy to escape the gravitational potential of MS~1512$-$cB58
($M_{\rm baryons} \sim 10^{10}$ M$_\odot$), but the large $\alpha$/Fe
ratio in the ISM of this object suggests that at least some of the
material made by previous stellar generations is retained.

The properties of the outflow in MS~1512$-$cB58 seem to be typical of
those in LBGs, although the data on fainter objects are necessarily
more uncertain (Pettini et al. 2001; Adelberger et al. 2003; Shapley
et al. 2003).  Large velocity shifts averaging $\sim$ 650 km~s$^{-1}$
are found by Shapley et al. (2003) between the low-ionization lines
and the Ly$\alpha$ emission lines of LBGs. A comparison with the study
of Pettini et al. (2001) suggests that these velocity offsets reduce to
$\sim$ 300 km~s$^{-1}$ when measured relative to the more reliable
nebular lines. This value is slightly higher than the outflow
velocities found in low-$z$ galaxies of similar SFR's (\S 4.4).

The equivalent widths of blueshifted low-ionization interstellar
absorption lines in LBGs anti-correlate strongly with the strength of
the Ly$\alpha$ emission, and correlate with the color excess
$E(B-V)$. The latter suggests that dust is entrained with the
outflowing neutral gas and is sufficiently widespread to redden the
host LBG; this effect is also observed in local wind galaxies
(\S4.10). The former is easy to understand if we recall that the
equivalent widths of saturated absorption lines (as is the case here)
depend only on the covering fraction of the absorbing material and the
range of velocities over which this material is absorbing.  Larger
covering factor or absorbing velocity interval implies that less
Ly$\alpha$ emission can escape from the wind. This effect may also
explain the redder UV continuum and the larger kinematic offsets
between Ly$\alpha$ and the interstellar absorption lines observed in
weak Ly$\alpha$-emitting LBGs (Shapley et al. 2003).

Quantifying the environmental impact of LBG winds comes by probing the
environment with spectra of background QSOs.  Around six $z \sim3$
LBGs, Adelberger et al. (2003) find hints of an H~I deficit within
comoving radius 0.5 $h_{70}^{-1}$ Mpc.  They favor a scenario whereby
LBG winds influence the nearby IGM directly over a proximity effect
caused by LBG-ionizing radiation.
The excess of absorption-line systems with large C~IV columns that
they find near LBGs is interpreted as further evidence for chemical
enrichment of the IGM by LBG winds, although it could also be
attributed to debris from tidal interactions (e.g., Morris \& van den
Berg 1994).

In a recent extension to this survey, Adelberger et al. (2005) find
that the relationship between C~IV column density and clustering
around LBGs is still present, although at a lower level than in the
original survey. Most galaxies now appear to have significant H~I
within 1 $h^{-1}$ Mpc of their centers, but some fraction of them
($\sim$ 1/3) still show a significant deficit. The LBGs with strong
H~I absorption have roughly as much absorption as expected from
windless, smoothed particle hydrodynamics (SPH) simulations of
$\Lambda$CDM universes (e.g., Croft et al. 2002; Kollmeier et al.
2003), but LBGs without much H~I are not present in these simulations.
These results are at odds with the original suggestion of large,
spherical windblown cavities around LBGs (Adelberger et al. 2003).
They are qualitatively more consistent with the idea that winds emerge
along paths of least resistance, possibly avoiding large-scale
filaments (e.g., Theuns et al. 2002).

\subsection{QSO Absorption-Line Systems}

Large-scale GWs at all redshifts can be detected by their absorption
of background QSO continuum light (e.g., Rauch 1998).
Comparisons between the number densities of Mg II absorbers and
star-forming galaxies and the properties of local outflows (e.g., Bond
et al. 2001) suggest that there are enough Mg II absorbers to account
for the expected properties of winds at $1 < z < 2$. But a more
detailed analysis of absorption-line spectra is needed to answer this
question quantitatively.  Possible absorption signatures of winds or
superbubbles include broad (few hundred km s$^{-1}$) and complex
profiles pointing to strong nonrotational kinematics, a pairwise
absorption pattern that straddles weak absorption near the kinematic
center of the line, $\alpha$-rich abundances, and large cloud-to-cloud
variations in metallicity and ionization level.  Such signals are
evident in several very strong Mg II absorbers (Bond et al. 2001) and
at lower column densities (e.g., Rauch et al. 2002; Zonak et
al. 2004).

Detailed comparison of the kinematics of absorbers with those of the
Mg II-absorbing galaxies can further test for winds. In a study of
five Mg II-absorbing galaxies at redshifts 0.44 $ < z <$ 0.66 and of
absorbers with projected impact parameters from those galaxies of 15
-- 75 $h^{-1}$ kpc, Steidel et al. (2002) find that halo rotation
sometimes dominates radial infall or outflow even for gas far from the
galactic plane.  But in the $z = 0.7450$ Mg II galaxy toward quasar
Q1331+17, no absorbing gas is detected at the projected velocity of
the disk rotation (Ellison, Mall\'en-Ornelas, \& Sawicki 2003).  The
motion of the absorbing gas in this galaxy is consistent with a large
($\sim$ 30 $h_{70}^{-1}$ kpc) superbubble expanding at $\sim \pm 75$
km~s$^{-1}$. Clearly, more data of high quality are needed for a final
verdict (see, e.g., C\^ot\'e et al. 2005).

\section{THE SIGNIFICANCE OF GALACTIC WINDS}

There is growing evidence that GWs have inhibited early star formation
and have ejected a significant fraction of the baryons once found in
galaxies.  The latter may explain why few baryons are in stars
($\Omega_*/\Omega_b \sim 0.1$; Fukugita, Hogan, \& Peebles 1998) and
why galaxies like the Milky Way contain fewer than expected from
hydrodynamical simulations (Silk 2003). We review in this section the
impact of winds on galaxies and on their environment.

\subsection{Influence of Winds on Galactic Scales}

\subsubsection{GALAXY LUMINOSITY FUNCTION}

GWs have modified substantially the shape of the galaxy luminosity
function, flattening its faint-end slope compared to that of the halo
mass function (e.g., Dekel \& Silk 1986; Somerville \& Primack 1999;
Benson et al. 2003; Dekel \& Woo 2003). 
The shallow potential of dwarf galaxies makes them vulnerable to
photoevaporation (if $v_c \la 10 - 15$ km s$^{-1}$; Barkana \& Loeb
1999), mechanical feedback (e.g., de Young \& Heckman 1994; MacLow \&
Ferrara 1999; Ferrara \& Tolstoy 2000), and ablation by GWs from
nearby galaxies (Scannapieco, Thacker, \& Davis 2001).
Significant feedback also appears necessary to avoid the `cooling
catastrophe' at high redshift that would otherwise overproduce massive
luminous galaxies (e.g., Cole et al. 2000).  Energies of a few
$\times~10^{49}$ ergs per solar mass of stars formed can explain the
sharp cutoff at the bright end of the luminosity function (Benson et
al. 2003). Starburst-driven winds are too feeble by a factor of
several to fully account for the cutoff. Benson et al. (2003)
therefore argue that feedback from BH accretion is the only way to
expel winds hot enough to prevent subsequent gas recapture by group
halos. The kinetic power supplied by jets in radio-loud AGN, $Q \sim
0.1~L_{\rm Edd}$ (see \S 5.4), may indeed suffice to account for the
paucity of high-mass systems. Feedback from starburst- and AGN-driven
winds may help set up the bi-modality observed in galaxy properties
(see below; also Dekel \& Birnboim 2005). AGN feedback may be
particularly effective in clustered environment where the infalling
gas is heated by a virial shock and thus more dilute.

\subsubsection{CHEMICAL EVOLUTION}

In the GW scenario, massive galaxies with deep gravitational
potentials are expected to retain more of their SN ejecta than dwarf
galaxies (e.g., Larson 1974; Wyse \& Silk 1985; Dekel \& Silk 1986;
Vader 1986).
Several authors have provided observational support for this picture,
often using luminosity as a surrogate for mass (e.g., 
Bender, Burstein, \& Faber 1993; Zaritsky et al. 1994; Jablonka,
Martin, \& Arimoto 1996; 
Trager et al. 1998; Kobulnicky \& Zaritsky 1999; Salzer et al. 2005).
An analysis of the Sloan Digital Sky Survey (SDSS) database by
Tremonti et al. (2004) has shown that the gas-phase metallicity of
local star-forming galaxies increases steeply with stellar mass from
10$^{8.5}$ to 10$^{10.5}$ M$_\odot h^{-2}_{70}$,
but flattens above 10$^{10.5}$ M$_\odot h^{-2}_{70}$. Similar trends
are seen when internal velocity or surface brightness is considered
instead of stellar mass (Kauffmann et al. 2003). The stellar mass
scale of this flattening coincides roughly with the dynamical mass
scale of metal retention derived by Garnett (2002).  In that paper,
Garnett used a simple closed-box chemical model to infer that the
effective yield increases with galaxy mass up to the stellar yield
obtained at $v_c \approx 125$ km s$^{-1}$.  These results suggest that
the chemical evolution of galaxies with $v_c \ga 125$ km s$^{-1}$ is
unaffected by GWs, whereas galaxies below this threshold tend to lose
a large fraction of their SN ejecta.  This is consistent with
estimates based on X-ray temperatures in wind galaxies (Sections 4.4,
4.6, and 4.9).

\subsubsection{DISK SIZE AND DARK MATTER CONCENTRATION}

Hydrodynamical simulations reveal that dynamical friction from the
inner dark halo acting on baryonic clumps overcompresses the galactic
disk by a factor of five in radius compared what is observed (e.g.,
Steinmetz \& Navarro 1999; Bullock et al. 2001).  The suppression by
stellar or AGN feedback of early gas cooling solves only part of this
angular momentum problem (e.g., Sommer-Larsen, G\"otz, \& Portinari
2003; Abadi et al. 2003).  Entrainment and removal of material with
low specific angular momentum by starburst- or AGN-driven GWs (e.g.,
Binney, Gerhard, \& Silk 2001; Maller \& Dekel 2002; Read \& Gilmore
2005) is needed to explain the formation of exponential disks and the
origin of bulge-less galaxies, especially if many bulges form by
secular evolution (Kormendy \& Kennicutt 2004).  If many of the
central baryons are thus ejected, CDM halo profiles become less cuspy
(e.g., Navarro, Eke, \& Frenk 1996) and more consistent with mass
distributions observed in dwarfs and low-surface brightness galaxies
(e.g., van den Bosch \& Swaters 2001; de Blok, McGaugh, \& Rubin 2001;
Gentile et al. 2004; see Ricotti \& Wilkinson 2004 for an alternative
viewpoint).

It is interesting to speculate which of the structural
properties observed in galaxies are due to the action of winds in
the early universe. For example, it is well known that the average
baryon density observed in a galaxy decreases with declining galaxy
luminosity or mass (e.g. Kormendy 1985). As we have mentioned, it 
is probable that powerful central winds in the early galaxy can
reshape the baryon density distribution in galaxies (e.g. van den Bosch
2001). This idea has been explored in many papers which incorporate
centrally driven feedback processes in order to patch up the
shortcomings of CDM simulations.
Similarly, there is a decreasing baryon/dark matter fraction observed
in galaxies with declining galaxy luminosity or mass (e.g. Persic,
Salucci \& Stel 1996; Mateo 1998). In a seminal paper, Dekel \&
Silk (1986) anticipated this now well-established trend after
considering the impact of SN-driven outflows in protodwarfs
(see also Nulsen \& Fabian 1997).

But there are reasons for believing that most of the feedback
prescriptions imposed in CDM hydro simulations to date have little
or no relevance to real galaxies (see Springel \& Hernquist 2003
for a recent summary); many of the prescriptions manifestly do not
conserve energy or entropy in the flow. When one considers the
evolution of GWs in different environments, the trends (i.e. those
arising from the action of winds) with total galaxy mass may be far
less marked.  For example, gas accretion onto a galaxy can staunch
the developing outflow for any galaxy mass (e.g. Fujita et al. 2004;
Springel \& Hernquist 2003). Therefore, it is not obvious which of
the structural properties can be ascribed to the action of winds at 
the present time.

\subsubsection{POROSITY OF HOST ISM}

The relative contribution of AGN and starbursts to the inferred
ionizing background depends critically on $f_{esc}$, the fraction of
ionizing photons that escape from each object.  The column densities
of disks imply $\tau \approx 10^4 - 10^7$ at the Lyman edge, so
leakage of ionizing radiation must be set by the topology of the
ISM. GWs should play a key role in clearing a path for the escaping
radiation (e.g., Dove, Shull, \& Ferrara 2000), but this has not yet
been confirmed observationally from constraints on
$f_{esc}$. H$\alpha$ measurements of high-velocity clouds above the
disk of our Galaxy indicate that the escape fraction normal to the
disk is 6\% ($f_{esc} \approx$ 1\% -- 2\% averaged over $4 \pi$ sr;
Bland-Hawthorn \& Maloney 1999, 2002). Estimates of the escape
fraction in local, UV-bright starburst galaxies yield $f_{esc} \le$
6\% (Heckman et al. 2001b and references therein), and similar values
are inferred for bright blue galaxies at $z = 1.1 - 1.4$ (Malkan,
Webb, \& Konopacky 2003). Star-forming galaxies thus contribute little
($<$ 15\%) to the ionizing background at $z \la 1.5$.  The situation
may be different at $z \ga 3$, where the comoving number density of
QSOs declines rapidly.
Steidel, Pettini, \& Adelberger (2001) infer $f_{esc} \approx 50
- 100\%$ for $\langle z \rangle = 3.4$ LBGs,
but these results have been questioned by Giallongo et al. (2002),
Fern\'andez-Soto, Lanzetta, \& Chen (2003), and Inoue et
al. (2005). The dark cores of the saturated interstellar absorption
lines in LBGs (Shapley et al. 2003) also appear inconsistent with
large $f_{esc}$, unless we see little of the escaping ionizing
radiation.  The large value of $f_{esc}$ inferred by Steidel et al. (2001),
if confirmed, may be due to powerful GWs in these objects (\S 6.1).

\subsubsection{SPHEROID -- BLACK HOLE CONNECTION}

The masses of the central BHs in early-type galaxies and bulges
correlate well with the velocity dispersions of the spheroidal
component: $M_{\rm BH} = 1.3 \times 10^8 \sigma_{200}^4$ M$_\odot$
(Ferrarese \& Merrit 2000; Gebhardt et al. 2000; Tremaine et
al. 2002). This correlation is remarkably similar to the Faber--Jackson
relation (Bernardi et al. 2003), and suggests a causal connection
between galaxy formation and BH growth by means of a GW that regulates
BH fueling (e.g., Silk \& Rees 1998; Haehnelt et al. 1998;
Fabian 1999; King 2003; Murray, Quataert, \& Thompson 2005; Begelman \& Nath 2005).
The wind may be produced by the starburst that accompanied the
formation of the spheroid or by the BH itself. An Eddington-like
luminosity is derived for the starburst or the BH, above which the
growth of both spheroid and BH is stopped by the wind. In the case of
a dominant BH wind, the Salpeter timescale, {\em i.e.} the timescale
for $M_{\rm BH}$ to double, must be similar to the star formation
timescale so that sufficient stars are formed before the BH wind blows
away the ambient gas and stops star formation (Murray et
al. 2005). The massive winds detected in nearby ULIRGs (\S 4.5) may be
local examples of what might have occurred as spheroids formed (e.g.,
high-$z$ LBGs and submm galaxies; \S 6.1).

\subsection{Influence of Winds on Intergalactic Scales}

\subsubsection{INTRACLUSTER MEDIUM}

Galaxy clusters are excellent laboratories to study the impact of GWs
on the environment because the hot, metal-enriched material ejected
from SNe is retained by the cluster gravitational potential. Most
metals in clusters are in the $\sim$ 0.3 $Z_\odot$ ICM, not in
galaxies. Several lines of evidence suggest that GWs, not ram-pressure
stripping, has dominated the transfer of metals from galaxies to ICM
(see review by Renzini 2004).  One is that
ejection of hot gas from proto-galaxy GWs can create the `entropy
floor'
(Kaiser 1991; Evrard \& Henry 1991) necessary to explain the steep
X-ray luminosity -- temperature relation for nearby groups and
clusters (e.g., Arnaud \& Evrard 1999; Helsdon \& Ponman 2000), the
lack of cluster evolution out to $z \sim 1$ (e.g., Mushotzky \& Scharf
1997), and the shallow density profiles of cooler groups (e.g., Horner
et al. 1999). Heating of $\sim$ 1 keV per gas particle would reproduce
these results.  Type II and Type Ia SNe (e.g., Lloyd-Davies, Ponman,
\& Cannon 2000), AGN (e.g.,
Cavaliere, Lapi, \& Menci 2002), and Type II SNe from very massive,
metal-poor progenitors ({\em i.e.}  Population III stars; e.g.,
Loewenstein 2001) may all contribute to the heating.

Analyses of ICM abundances provide some constraints on the relative
importance of these energy sources. Early reports of large
$\alpha$-element abundances in bright clusters by Mushotzky et
al. (1996)
first showed that Type II SNe enrich (and thus heat) some of the
ICM. ASCA and XMM results now suggest that iron-rich Type Ia
ejecta dominate in the centers of rich clusters, whereas the
$\alpha$-rich products of Type II SNe are distributed more evenly
(e.g., Finoguenov et al. 2002, Tamura et al. 2004 and references
therein).  The iron mass scales with the optical light from the
early-type galaxies and the cluster X-ray luminosity (e.g., Arnaud et
al. 1992; de~Grandi et al. 2004), suggesting iron enrichment
by Type Ia SNe from these galaxies. A contribution from Population
III stars may be needed to explain the inhomogeneity of
$\alpha$-elements in the ICM (Baumgartner et al. 2005).
In-situ enrichment by intracluster stars may also be significant
(Zaritsky, Gonzalez, \& Zabludoff 2004).

AGN winds help enrich the ICM with metals, and the ubiquity of large
``cavities'' in the X-ray surface brightness of
clusters with radio galaxies (e.g.,
B\"ohringer et al. 1993; Fabian et al. 2000; McNamara et al. 2000, 2001;
Heinz et al. 2002; Mathews \& Brighenti 2003)
confirms that they modify the thermodynamics of the ICM.  The hot,
relativistic gas injected into the ICM by the AGN reduces, and perhaps
even quenches, the mass accretion of cooling flows. The exact
mechanism by which energy in the radio bubbles turns into heat is
still debated, but the absence of strong shocks along cavity walls,
and the discovery of low-amplitude, semi-periodic ripples in the
Perseus cluster (Fabian et al. 2003) suggest that viscous dissipation
of sound waves may heat much of the inner ICM (see also Ruszkowski,
Br\"uggen, \& Begelman 2004a, 2004b; Reynolds et al. 2005). Other
possible heaters include thermal conduction and turbulent mixing
(e.g., Narayan \& Medvedev 2001;
Ruzkowski \& Begelman 2002; Kim \& Narayan 2003a, 2003b).

\subsubsection{INTERGALACTIC MEDIUM}

The sphere of influence of GWs appears to extend to the low-density
environment of the Ly$\alpha$ forest [N(H~I) $\la$ 10$^{17}$
cm$^{-2}$]. Here, metallicities of 0.1\% -- 1\% solar have been
measured, with a possible excess of $\alpha$-rich SN II products in
the denser clouds (e.g., Rauch, Haehnelt, \& Steinmetz 1997; Songaila
1997; Hellsten et al. 1997; Dav\'e et al. 1998; Carswell, Shaye, \&
Kim 2002).
The detection of metals in the IGM seems to favor momentum- over
energy-driven winds (\S 2.3), or scenarios where the winds emerge
along paths of least resistance without disturbing the filaments
responsible for the Ly$\alpha$ forest (e.g., Theuns et al. 2002).
Remarkably, both the column density distribution of C~IV absorbers and
its integral ($\Omega_{C~IV}$) are invariant over $2 \la z \la 5$
(Songaila 2001; Pettini et al. 2003). One possible explanation is that
most of the IGM metals are already in place by $z \sim 5$, perhaps
from SN-driven outflows from low-mass subgalactic systems (e.g., Qian
\& Wasserburg 2005). Such systems may also be responsible for
reionizing the IGM (Loeb \& Barkana 2001 and references
therein). However, this scenario does not completely explain why
$\Omega_{\rm C~IV}$ remains constant over this redshift range despite
variations in the intensity and spectrum of the ionizing background
(\S 7.1.4). Alternatively, the C~IV systems are associated directly
with GWs from LBGs at $z \la 5$, and the constancy of $\Omega_{\rm C~IV}$
arises instead from the flatness of the SFR density over $z \simeq 1.5
- 4$ (Adelberger et al. 2003).  A critical discriminator between these
two scenarios is to measure the metallicity in truly intergalactic
clouds with N(H~I) $\la$ 10$^{14}$ cm$^{-2}$ (Cen, Nagamine, \&
Ostriker 2005). This is a portion of the Ly$\alpha$ forest that has
not yet been explored in detail (although see Ellison et al. 2000).

\section{FUTURE DIRECTIONS}
    
Although great strides have been made over the past 25 years in
understanding the physics and impact of GWs in the local and distant
universe, much work remains to be done to quantify the role of these
winds on the formation and evolution of galaxy-sized structures and
the intergalactic environment. We now outline observational and
theoretical issues that we feel deserve urgent attention.

\subsection{Observational Challenges}

\subsubsection{UNBIASED CENSUS OF LOCAL WINDS}
As noted in \S 4.1, our current sample of GWs detects outflows that
are sufficiently large and/or powerful but not so energetic as to
expel all gas from their hosts. For example, the Galaxy's wind (\S
3.2) would be undetectable beyond the Local Group.
While it will be difficult to detect blown-away relics, there is a
clear need to search the local volume systematically for winds.
At optical wavelengths, the advent of tunable filters on 8-meter class
telescopes will improve tenfold the sensitivity of optical wind
surveys.  These instruments will be ideal for searching for galaxies
with starburst-driven winds through the contrast in gaseous excitation
between wind and star-forming disk (\S 4.8).
An IF spectrometer equipped with adaptive optics would complement
tunable filters by providing densely sampled data on kinematics,
filling factor, and excitation processes. CXO and XMM-Newton will
continue to harvest high-quality data on the hot medium in GWs. New
long-wavelength radio telescopes (e.g., GMRT, GBT, EVLA, and
SKA) can better search for the relativistic component of GWs.
Particularly important will be to determine the relative importance of
GWs in dwarf and massive galaxies (\S 4.5).

\subsubsection{WIND FLUID}

This component drives starburst-driven winds, yet has been detected in
very few objects.  Metal abundances suggest enrichment by SNe II, but
the measurements are highly uncertain. Both sensitivity and high
spatial resolution are needed to isolate the hot wind fluid from X-ray
stellar binaries and the rest of the X-ray-emitting gas.  But, no such
instrument is planned for the foreseeable future.  Indirect methods
that rely on the properties of gas in the energy injection zone to
constrain the wind pressure may be necessary. Current measurements of
the pressure profiles in wind galaxies are certainly contaminated by
the foreground/background disk ISM. Measurements in the mid- or far-IR
with the {\em Spitzer Space Telescope} (SST) and {\em Herschel Space
  Observatory} will reduce the effects of dust obscuration.

\subsubsection{ENTRAINED MOLECULAR GAS \& DUST}

Despite the important role of the molecular component in GWs,
high-quality mm-wave data exist only for M82. This is due to the
limited sensitivity and spatial resolution of current instruments, but
this will change soon.  New mm-wave arrays (e.g., CARMA, and
especially ALMA) will map the molecular gas in a large sample of
nearby galaxies with excellent resolution ($<$ 1\arcsec). Sub-mm and
mid-IR data from the ground (e.g., SMA, JCMT, CSO) and from space
(e.g., SST and Herschel) will constrain the amount and location of
dust in the winds.

\subsubsection{ZONE OF INFLUENCE \& ESCAPE EFFICIENCY}

The environmental impact of GWs depends on the size of their zone of
influence and on the fraction of wind fluid and entrained ISM that can
vent from their hosts. Very deep emission-line, X-ray, and radio data
on large scale would help tremendously to constrain wind
extent. Tunable filters on 8-meter class telescopes may be
particularly useful here.  Absorption-line studies of bright
background galaxies (e.g., high-$z$ quasars, LBGs) have proven to be a
very powerful tool to constrain the zone of influence of GWs at large
redshifts. The Cosmic Origins Spectrograph (COS) on HST
could extend the sample to a larger set of wind galaxies.  Deep 21-cm
maps of GW hosts on scales of up to $\sim$ 100 kpc would help to
quantify the effects of halo drag.  The escape efficiency of winds may
also be constrained indirectly by measuring the stellar metallicities
of galaxies suspected to have experienced GWs (e.g., largely gas-free
dwarf spheroids in the Local Group) and then comparing these values
with the predictions of leaky-box models (e.g., Lanfrancini \&
Matteucci 2004).

\subsubsection{THERMALIZATION EFFICIENCY}

Observational constraints on the thermalization efficiency of GWs are
rare because of an incomplete accounting of the various sources of
thermal energy and KE in the wind.  A multiwavelength approach that
considers all gas phases is needed.

\subsubsection{WIND/ISM INTERFACE \& MAGNETIC FIELDS}

Constraints on microphysics at the interface between the wind and
galaxy ISM are available in only a handful of
galaxies. High-resolution ($\la$ parsec scale) imaging and spectra of the
entrained disk material in a sizable sample of local objects are
required. The large-scale morphology of the magnetic field lines has
been mapped in a few winds, but the strength of the field on pc scale
is unknown.  This information is crucial in estimating the
conductivity between the hot and cold fluids.

\subsubsection{GALACTIC WINDS IN THE DISTANT UNIVERSE}

Absorption-line studies of high-$z$ galaxies and QSOs will remain a
powerful tool to search for distant GWs and to constrain their
environmental impact.  Future large ground and space telescopes will
extend such studies to the reionization epoch.  These galaxies are
very faint, but gravitational lensing by foreground clusters can make
them detectable and even spatially resolved. Cross-correlation
analyses of wind galaxy surveys with detailed maps of the cosmic
background radiation (CBR) (e.g., from the Planck mission) may also
help to constrain the extent of the hot medium in winds by means of
the Sunyaev-Zel'dovich effect (e.g., Voit 1994; Scannapieco \&
Broadhurst 2001), although one will need to consider all other
foreground sources that affect the CBR
(Hern\`andez-Monteagudo, Genova-Santos, \& Atrio-Barandela 2004 and
references therein).

\subsubsection{POSITIVE FEEDBACK BY WINDS}

Star-forming radio jet/gas interactions have been found in a few
nearby systems (e.g., Minkowski's Object: van Breugel et al. 1985;
Cen~A: Oosterloo \& Morganti 2005) and are suspected to be responsible
for the ``alignment effect'' between the radio and UV continua in
distant radio galaxies (e.g., van Breugel et al. 2004 and references
therein). The same physics may also provide positive feedback in wind
galaxies.  Convincing evidence for superbubble-induced star formation
has recently been found in the disk of our own Galaxy (Oey et
al. 2005). In \S 4.3 we noted that shocked H$_2$ gas and circumnuclear
rings of H~II regions in a few wind galaxies may represent
wind-induced star formation at the contact discontinuity/ISM shock
associated with lateral stagnation of the wind in the galaxy disk.
This region is also a gas reservoir from which to fuel the starburst.
We do not know how often such rings form.  Excess free-free emission
on the inner edge of the outflow near the disk in M82 has been
interpreted as a wind-induced starburst (Matsushita et al. 2004). A
galaxy companion within the zone of influence of a GW may also be
searched for wind-induced starburst activity (e.g., Irwin et
al. 1987).  This effect may have triggered the starburst in NGC~3073,
a companion to NGC~3079 (e.g., Filippenko \& Sargent 1992).

\subsection{Theoretical Challenges}

\subsubsection{MODELING THE ENERGY SOURCE}

Current simulations do a poor job of modeling the
energy source itself, especially in AGN-driven winds
where energy and momentum injection rates are virtually unknown.
Deeper understanding of AGN jets and winds is needed before simulating
the impact of AGN-driven outflows.  The situation for starburst-driven
winds is much better, but the input energetics are still highly
uncertain because the thermalization efficiency is constrained poorly
by observation and theory.  Simulations by Thornton et al. (1998) have
shown that radiative losses of SN remnants expanding into {\em
uniform} media of $\sim$ 0.02 -- 10 cm$^{-3}$ are $\sim$ 0\% --
90\%. But, it would be useful to run more realistic simulations with a
range of molecular filling factors for young star clusters that evolve
within a multiphase ISM.

\subsubsection{MODELING THE HOST ISM}

The work of Sutherland et al. (2003b) noted in \S 2.4 is the first of
a new generation of simulations able to handle a multiphase ISM with a
broad range of densities and temperatures.  Such sophistication is
crucial to understanding and predicting the mass of gas entrained in
winds.
As discussed in \S 2.4, simulations show that the initial encounter of
clouds with a wind drives a strong shock that may devastate the
clouds.  Once in ram-pressure equilibrium with the wind, however,
clouds may accelerate to a significant fraction of the wind velocity
before RT and KH instabilities shred them. To test the survival of
entrained gas, these hydro processes should be combined with the
effects of conductive evaporation to model the interface between the
hot wind fluid and the dense ISM clouds.  On the large scale, it will
be important to use realistic distributions for the galaxy ISM,
accounting for the clumpiness of the halo component and the disk
(e.g., Fig. 1), and possible large-scale magnetic fields.  These
simulations would quantify the drag of the halo gas, the impact of
wind on disk ISM, and the feedback from wind-induced star formation.

\subsubsection{COUPLING THE RADIATION FIELD TO GAS}

Current simulations do not account for possible coupling between the
wind material and the radiation field emitted by the energy source or
the wind itself, and indeed ignore radiation pressure.  For instance,
thick, dusty ISM clouds entrained in the wind may be photoionized by
the hot wind fluid and radiative shocks at the wind/ISM interface,
with major impact on their gaseous ionization.

\medskip
\noindent{\em Acknowledgements}
This article was begun while S.V.\ was on sabbatical at the California
Institute of Technology and the Observatories of the Carnegie
Institution of Washington; this author thanks both institutions for
their hospitality.  G.C. thanks CTIO for its hospitality. We thank the
scientific editor, J. Kormendy, for constructive comments on the style
and contents of this review, and S. Aalto, P.-A. Duc, R. Braun, D.
Forbes, L. Tacconi, D. Calzetti for organizing recent conferences that
provided excellent venues to discuss many of the issues reviewed here.
S.V. acknowledges partial support of this research by
NSF/CAREER grant AST-9874973.

\twocolumn

\end{document}